\documentclass[12pt]{JHEP3}
\usepackage{graphicx}
\usepackage{amssymb}
\usepackage{amsmath}
\usepackage{epsfig}
\usepackage{dcolumn}

\title{Thermodynamics and Cosmological Constant of Non-Local Field Theories from p-Adic Strings}

\author{Tirthabir Biswas\\
Department of Physics\\
St. Cloud State University, St. Cloud, MN 56301\\
E-mail: \email{tbiswas@gravity.psu.edu}}

\author{Jose A. R. Cembranos\\
William I. Fine Theoretical Physics Institute\\
University of Minnesota, Minneapolis, MN 55455\\
E-mail: \email{cembranos@physics.umn.edu}}

\author{Joseph I. Kapusta\\
School of Physics and Astronomy\\
University of Minnesota, Minneapolis, MN 55455\\
E-mail: \email{kapusta@physics.umn.edu}}

\date{\today}

\abstract{We develop the thermodynamics of field theories
characterized by non-local propagators. We analyze the partition
function and main thermodynamic properties arising  from
perturbative thermal loops. We focus on the p-adic models associated
with the tachyon phenomenology in string theories. We reproduce well
known features of these theories, but also obtain many new results. In
particular, we explain how to maintain consistency of such non-local
theories by avoiding the appearance of ghosts at finite temperature.
As a consequence of this fact, the vacuum energy in p-adic theories
becomes positive. It is also hierarchically suppressed, and we explore the
parameter space where it is consistent with the observed value of
the cosmological constant.}


\def\thalf{{\textstyle{\frac{1}{2}}}}

\newcommand{\be}{\begin{equation}}
\newcommand{\ee}{\end{equation}}
\newcommand{\ba}{\begin{eqnarray}}
\newcommand{\ea}{\end{eqnarray}}
\newcommand{\bd}{\begin{displaymath}}
\newcommand{\ed}{\end{displaymath}}

\newcommand{\bt}{\beta}
\newcommand{\ga}{\gamma}

\newcommand{\da}{\delta}
\newcommand{\la}{\lambda}
\newcommand{\za}{\zeta}

\newcommand{\en}{\epsilon}
\newcommand{\oa}{\omega}
\newcommand{\Ga}{\Gamma}

\newcommand{\La}{\Lambda}

\newcommand{\cR}{{\cal R}}
\newcommand{\cO}{{\cal O}}

\newcommand{\n}{\nabla}

\newcommand{\ra}{\rightarrow}

\newcommand{\LF}{\left(}
\newcommand{\RF}{\right)}
\newcommand{\LT}{\left[}
\newcommand{\RT}{\right]}

\newcommand{\Rd}{\right.}

\newcommand{\mx}{\mbox}

\newcommand{\for}{\mx{ for }}

\newcommand{\with}{\mx{ with }}

\newcommand{\ie}{{\it i.e.\ }}

\newcommand{\non}{\nonumber}
\newcommand{\Sf}{\varsigma}
\newcommand{\Ff}{\chi}

\newcommand{\ms}{m_s}
\newcommand{\Dn}{d}

\begin{document}

\maketitle

\begin{center}

\vspace{1cm}

\end{center}

\newpage

\section{Introduction}

In this paper we study a special class of field theories which have an infinite number of higher derivative terms in the form of an exponential. Such non-local structure of quantum field theories is recurrent in many stringy models. For instance, this is the case for tachyonic actions in string field theory \cite{sft1}-\cite{pressure} (for a review see \cite{sft_review}), bulk fields localized on codimension-2 branes \cite{claudia} and various toy models of string theory such as $p$-adic strings \cite{padic_st,zwiebach}, zeta strings \cite{zeta}, and strings quantized on a random lattice
\cite{random}-\cite{ghoshal}. More general non-local theories, in which the derivatives do not necessarily appear in the combination $\Box = -\partial_t^2 + \n^2$, arise in noncommutative field theories \cite{ncft}, field theories with a minimal length scale \cite{minimal} (such as doubly special relativity), fluid dynamics \cite{kdv,fluid} and quantum algebras \cite{kdv}.

In addition, like most higher-derivative theories, these theories
have better ultra-violet (UV) convergence.  Unlike {\it
finite-order} higher derivative theories, by virtue of having an
infinite set of higher derivative terms, they have been conjectured
to be free of ghosts\footnote{There are no perturbative states or
poles in the propagator.  There are instabilities causing
oscillations to grow unboundedly \cite{zwiebach}, but this is due to
the presence of the tachyon \cite{futuret}.} and to have a
well-posed initial value problem \cite{math,zwiebach,neil,gianluca}
making them phenomenologically interesting to study. In particular,
these models have been found to provide novel cosmological
properties such as non-slow-roll inflation \cite{bbc}, crossing of
the phantom-divide in the context of dark energy \cite{cosmo},
non-singular bouncing solutions \cite{bouncing} (also see
\cite{warren} for similar work with non-local gravitational actions)
among others \cite{cosmology}. However, most of these analysis have
largely been classical, and little attention has been paid to
quantum loop calculations, but see \cite{marc,minahan} for an
exception.

Our goal here is to analyze the thermodynamic properties of this type of theory. There are three main motivations for this study.  Firstly, as was explained in \cite{prl-old}, some of the results in the $p$-adic theory closely parallel the ones that comes from purely stringy calculations\footnote{For a different field theoretic approach aiming to capture the physics of the Hagedorn phase, see \cite{seiberg}.}. The computations involved are relatively simple but we can nevertheless probe both high and low temperature regimes (with respect to the Hagedorn temperature), and therefore hope to clarify some aspects of string thermodynamics.

Secondly, it is well documented that the thermal history of our universe plays a crucial role in defining our cosmology. In particular there have been recent studies involving the Hagedorn phase in the early universe
\cite{vafa}-\cite{hagedorn-cosmology}, especially to see whether cosmological perturbations can have a thermal origin \cite{thermal,hag-bounce}. Our results may prove useful in this context.

Thirdly, higher derivative alternatives to supersymmetric extensions of the Standard Model have recently been proposed \cite{lee-wick}. In brief, these theories attempt to upgrade the mass scale in a Pauli-Villars regularization scheme to a physical parameter with observational consequences, possibly detectable at the Large Hadron Collider (LHC). Since the non-local modifications to the propagator in $p$-adic or string field theory models make the loop integrals finite, one could possibly treat the mass scale associated with the non-locality as a regularization parameter. On the other hand one can also try to explore whether similar consistent, both theoretical and phenomenological, extensions of the Standard Model exists if we believe the non-locality to be physical and possibly at the TeV scale; see \cite{moffat} in this direction. The tools that we have developed to deal with quantum corrections should prove helpful in this context. As an interesting by-product, while exploring quantum consistency of these models, we found that suitable insertions of counter-terms make the vacuum energy positive. Further, this energy can be hierarchically suppressed if the string scale small compared to the Planck scale, and therefore lends itself to the phenomenology of dark energy.

\begin{figure}[htbp]
\begin{center}
\includegraphics[width=0.55\textwidth,angle=0]{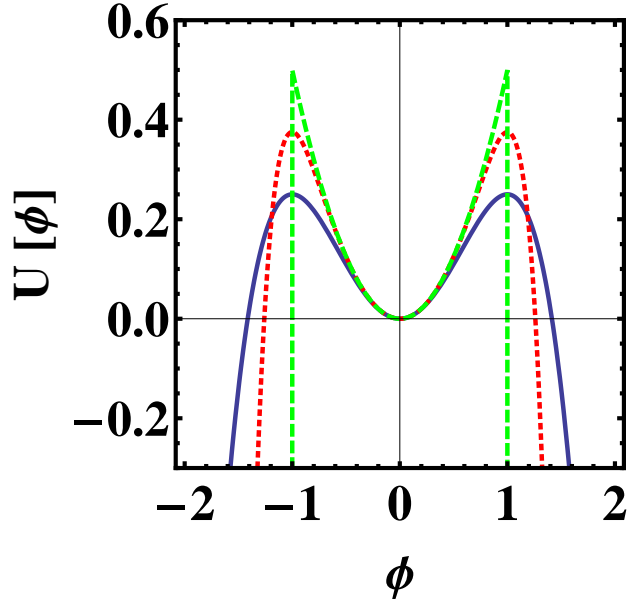}
\end{center}
\caption{The potential of the $p$-adic tachyon for $p$ = 3, 7 and $p \rightarrow \infty$.
\label{potfig}}
\end{figure}

In this paper our focus will be on the $p$-adic string theories, whose action is given by \cite{witten,Frampton}
\begin{equation}
S =
\frac{\ms^D}{g_p^2} \int d^D x  \left[ -\frac{1}{2} \varphi\,
{\rm e}^{-{\Box/ M^2}} \varphi+\frac{1}{p+1} \varphi^{p+1} \right]\,,
\label{action}
\end{equation}
where $\Box = -\partial_t^2 + \nabla_{D-1}^2$ in flat space, and we have defined
\be
\frac{1}{g_p^2} \equiv
\frac{1}{g_o^2}\frac{p^2}{p-1} \;\;\;\; {\rm and} \;\;\;\; M^2(p)
\equiv \frac{2\ms^2}{\ln p} \,.
\label{gpmp}
\ee
The dimensionless scalar field $\varphi(x)$ describes the open string tachyon, $\ms$ is the string mass scale, and $g_o$ is the open string coupling constant.  Though the action (\ref{action}) was originally derived for $p$ a prime number, it appears that it can be continued to any positive integer and even makes sense in the limit $p\rightarrow 1$ \cite{p=1}.  Setting $\Box=0$ in the action, the resulting potential takes the form $U = (\ms^D/g_p^2)(\frac12 \varphi^2 -
\frac{1}{p+1}\varphi^{p+1})$. Its shape is shown in Figure \ref{potfig}.

The action (\ref{action}) is a simplified model of the bosonic string which only qualitatively reproduces some aspects of a more realistic theory; that being said, there are several nontrivial similarities between $p$-adic string theory and the full string theory.  For example, near the true vacuum of the theory ($\varphi=0$) the field naively has no particle-like excitations since its mass squared goes to infinity\footnote{Reference \cite{zwiebach} found anharmonic oscillations around the vacuum by numerically solving the full nonlinear equation of motion. However, these solutions do not correspond to conventional particle states.}.  This is the $p$-adic version of the statement that there are no open string excitations of the tachyon vacuum.  A second similarity is the existence of lump-like soliton solutions representing $p$-adic D-branes
\cite{Sen:p-adic}.  The theory of small fluctuations about these lump solutions has a spectrum of equally spaced masses-squared for the modes \cite{Minahan:ModeInteractions}, as in the case of normal bosonic string theory. It is worth pointing out that one obtains a very similar action with exponential kinetic operators (and usually assumed to have a cubic or quartic potential) while quantizing strings on a random lattice \cite{random}.  These field theories are also known to reproduce several features, such as the Regge behavior \cite{marc}, of their stringy duals. Although our analysis focuses on the specific $p$-adic action, it can easily be applied to such theories as well.

In this present paper we develop the field theoretic tools to
compute thermodynamic quantities for such non-local models. In
\cite{prl-old} we had already explained how these models reproduce
some of the results of string theory. Here we provide details of
these calculations. Second, we point out how quantum consistency of
such models (that the theory remains ghost free) can be maintained
by inserting suitable counter-terms. Surprisingly, we find that our
prescription leads to some interesting consequences for the
cosmological constant. The cosmological constant can not only be
positive but it can be hierarchically suppressed if the string scale
is small compared to the Planck scale. Finally, we discuss
non-perturbative effects. This includes summing over classes of
infinite numbers of diagrams which gives rise to singularities at
non-zero coupling constants. In \cite{Biswas:2010yx} we will study
finite temperature solitonic solutions leading to contributions
which are divergent as the coupling vanishes.

The paper is organized as follows. In the next section we start with the finite temperature formulation of these non-local field theories with emphasis on the free theory. In section \ref{sec:2-loop} we focus on $\varphi^4$ interactions corresponding to $p=3$ and $D=4$. We compute the 2-loop partition function and briefly discuss the connections to string theory. In section
\ref{sec:self-energy} we focus on consistency of the model with respect to quantum loop corrections, and provide a prescription for counter-terms which needs to be added to address this issue. We also discuss the implication of our prescription for the cosmological constant.  In section \ref{sec:higher-loops} we focus on contributions to the free energy at arbitrary higher loops and comment on the validity of the perturbative analysis. In section \ref{sec:higher-p} we generalize our results to arbitrary values of $p$, while in section~\ref{sec:cc}, we study $p$-adic theories in arbitrary dimensions. The main aim in this section is to discuss its relevance for the cosmological constant.  Finally, in section \ref{sec:conclusion}, we conclude with a discussion on unresolved issues and allude to future research directions that may be able to shed some light on them.

\section{Free Energy at Zero Order: No Particle Degrees of Freedom}
\label{sec:1-loop}

Consider the action for $D=4$ space-time
dimensions and $p=3$. The finite temperature action corresponding to
(\ref{action}) is
\be S = \int_0^{\beta} d\tau \int d^3x \left[
-\thalf \phi({\bf x},\tau) {\rm e}^{-(\partial^2/\partial \tau^2 +
\nabla^2)/M^2} \phi({\bf x},\tau) - \lambda \phi^4({\bf x},\tau)
\right] \, , \label{3adic}
\ee
where we have performed the rescaling
\be
\phi\equiv{m_s^{2} \over g_3}\varphi \,,\  \la\equiv -{1\over
18} {g_o^2\over m_s^{4}}\,.
\ee
Notice that $\phi$ has dimension of energy-squared.  We could
multiply the kinetic term by some positive parameter with dimension
of energy-squared so that $\phi$ has dimension of energy, but it is
actually more convenient to write it this way.  Note that $\lambda$
has dimension of energy to the minus fourth power.  Analysis of this
model follows very closely that of the usual scalar theory at finite
temperature.  See Chapters 2 and 3 of Ref. \cite{KapGale}.

To perform the functional integral we use the Fourier transform
\be
\phi({\bf x},\tau) = \frac{1}{\sqrt{\beta V}} \sum_n \sum_{\bf k}
{\rm e}^{i({\bf k}\cdot{\bf x}+\omega_n \tau)} \phi_n({\bf k}) \, .
\ee
The Fourier amplitude $\phi_n({\bf k})$ is dimensionless, which is very convenient for performing functional integrals.  The Matsubara frequency is $\omega_n = 2\pi nT$.  After integration over space and imaginary time we get the free action
\be
S_0 = - \thalf \sum_n \sum_{\bf k} D_0^{-1}(\omega_n,{\bf k})
\phi_n^*({\bf k}) \phi_n({\bf k}) \, ,
\ee
where we have used $\phi_n^*({\bf k}) = \phi_{-n}({-\bf k})$; this is a consequence of the reality of the scalar field.  This action defines the free propagator
\be
D_0(\omega_n,{\bf k}) = {\rm e}^{-(\omega_n^2 + {\bf k}^2)/M^2} \, .
\ee

The partition function to zero order in $\lambda$ is
\ba
Z_0 &=& N' \prod_n \prod_{\bf k}
\left[ \int_{-\infty}^{\infty} dA_n({\bf k})
{\rm e}^{-\thalf D_0^{-1}(\omega_n,{\bf k}) A_n^2({\bf k})} \right]
 \nonumber \\
&=& N' \prod_n \prod_{\bf k} \left[ 2\pi D_0(\omega_n,{\bf k}) \right]^{1/2} \, .
\ea
Here $N'$ is an as yet arbitrary constant.  Integration is over the amplitudes of the Fourier components since the phase of $\phi_n({\bf k})$ drops out of the action.  Taking the logarithm
\be
\ln Z_0 = \ln N' + \thalf \ln(2\pi) \sum_n \sum_{\bf k}
+ \thalf \sum_n \sum_{\bf k} \ln [D_0(\omega_n,{\bf k})] \, .
\ee
As usual one chooses $N'$ such that the sum of the first two terms cancel.  Multiplication of the partition function by a temperature independent constant does not change the thermodynamics.  Thus
\be
\ln Z_0 = - \thalf \sum_n \sum_{\bf k} \frac{\omega_n^2 + {\bf k}^2}{M^2} \, .
\ee
We will now use two different approaches to show that this expression is zero.

In the first approach we express the sum as a contour integral.  The general formula is
\bd
T \sum_n f(k_0 = i \omega_n) = \frac{1}{4\pi i} \int_{-i\infty}^{i\infty} dk_0 \left[ f(k_0) + f(-k_0) \right]
\ed
\be
+ \frac{1}{2\pi i} \int_{-i\infty + \epsilon}^{i\infty+ \epsilon} dk_0 \left[ f(k_0) + f(-k_0) \right] \frac{1}{{\rm e}^{\beta k_0}-1} \, ,
\ee
under the assumption that the function $f(k_0)$ has no singularities on the imaginary axis.  The first integral on the right side is referred to as a vacuum contribution.  It can be converted to Euclidean space using $k_0=ik_4$.  Then the expression
\bd
\int \frac{dk_4 d^3k}{(2\pi)^4} \left( \frac{k_4^2 + {\bf k}^2}{M^2} \right)
\ed
is zero by the usual regularization procedure.  The second integral on the right side is referred to as a finite temperature contribution.  Since $f(k_0) =
(-k_0^2 + {\bf k}^2)/M^2$ is analytic in the right half plane, the contour can be pushed to infinity, showing that it is zero.

In the second approach we write the logarithm of the partition function as
\be
\ln Z_0 =-{V\over 2M^2}\sum_n\int d^3k\ (k^2+(2\pi nT)^2)
\ee
We use dimensional regularization to compute the above integral, and introduce two small parameters, $\en$ and $\da$.
\be
I_n(\en,\da)=\int {d^{3+\en}k\over (k^2+(2\pi nT)^2)^{-1+\da}}
\ee
We then use the standard formula
\be
\int \frac{d^Dp}{(2\pi)^D}\frac{k^A}{(k^2+m^2)^B}=\frac{\Ga(B-A-D/2)\Ga(A+D/2)m^{2A-2B+D}}{(4\pi)^{D/2}\Ga(B)\Ga(D/2)} \, .
\ee
Substituting $B=-1+\da$, $D=3+\en$, $A=0$ and $m=2\pi nT$ we have
\ba
I_n(\en,\da)&=&\frac{\Ga(-5/2+\da-\en/2)\Ga(3/2+\en/2)(2\pi nT)^{5-2\da+\en}}{(4\pi)^{3/2+\en/2}\Ga(-1+\da)\Ga(3/2+\en/2)} \nonumber \\
&=& \frac{\Ga(-5/2+\da-\en/2)(2\pi nT)^{5-2\da+\en}}{(4\pi)^{3/2+\en/2}\Ga(-1+\da)} \, .
\ea
We thus have
$$\ln Z_0=-{V\over 2M^2}\sum_n I_n=-{V\Ga(-5/2+\da-\en/2)(2\pi T)^{5-2\da+\en}\over 2M^2(4\pi)^{3/2+\en/2}\Ga(-1+\da)}\sum_nn^{5-2\da+\en}$$
$$=-{V\Ga(-5/2+\da-\en/2)(2\pi T)^{5-2\da+\en}\over M^2(4\pi)^{3/2+\en/2}\Ga(-1+\da)}\za(5-2\da+\en)
$$
where $\za$ is the Reimann zeta function. We now observe that as $\en$ and $\da\ra0$ all the terms except $\Ga(-1+\da)$ are regular. Thus using the Laurant series expansion
\be
\Ga(-n+\en)=\frac{(-1)^n}{n!}\LT \frac{1}{\en}-\ga+\psi(n)+\cO(\en)\RT
\ee
where $\ga\approx 0.5772$ is the Euler-Mascheroni constant, and
\be
\psi(n)=\sum_{i=1}^n{1\over i}\ ,
\ee
we find
\be
\ln Z_0 =  V\LT{2\pi^{7/2}\Ga(-5/2)\za(5)T^{5}\over M^2}\RT \da +\cO(\da^2,\en^2,\da\en)
\ee
Once again we find that $\ln Z_0$ vanishes.

The fact that the free theory gives no contribution to the pressure, energy density or entropy density should not be a surprise.  The free propagator has no poles and therefore no thermal excitations.  This is consistent with the fact that free $p$-adic theories are trivial in the sense that they contain no physical particle-like degrees of freedom.

\begin{figure}
\begin{center}
\includegraphics[width=2.5in,angle=0]{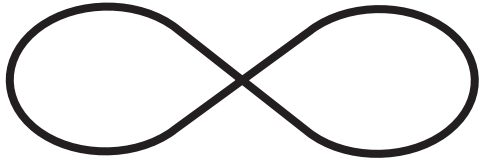}
\caption{This diagram constitutes the sole contribution to first order in $\lambda$ for the case $p=3$.}
\end{center}
\label{leadingp3}
\end{figure}

\section{Free Energy at First Order and Thermal Duality}
\label{sec:2-loop}

For interacting non-local theories such as
(\ref{3adic}), the Feynman rules are identical to those of the usual
scalar theory with the use of the appropriate propagator.  The only
diagram at first order in the coupling $\lambda$ is shown in Figure
\ref{leadingp3}.  There is a combinatoric factor of 3 and a factor
$(-\lambda)$ associated with the vertex. It leads to \be \ln Z_1 =
3(-\lambda) \beta V \left[ T \sum_n \int \frac{d^3k}{(2\pi)^3}
D_0(\omega_n,{\bf k}) \right]^2\,. \label{Z2loop} \ee Due to the
exponential nature of the bare propagator the loop diagrams are
expected to be convergent in both the IR and UV.  A useful formula
is
\be
\sum_n \int \frac{d^3k}{(2\pi)^3} D_0^N(\omega_n,{\bf k}) =
\left(\frac{M}{2\sqrt{N\pi}}\right)^3 \Sf\left( \frac{2\sqrt{N}\pi
T}{M}\right)\,, \label{useful1} \ee
where \be \Sf(x) =
\sum_{n=-\infty}^{\infty} {\rm e}^{-n^2x^2}=\theta_3(0,e^{-x^2})\ ,
\ee
the third Jacobi elliptic theta function.
\begin{figure}
\begin{center}
\includegraphics[width=3.2in,angle=0]{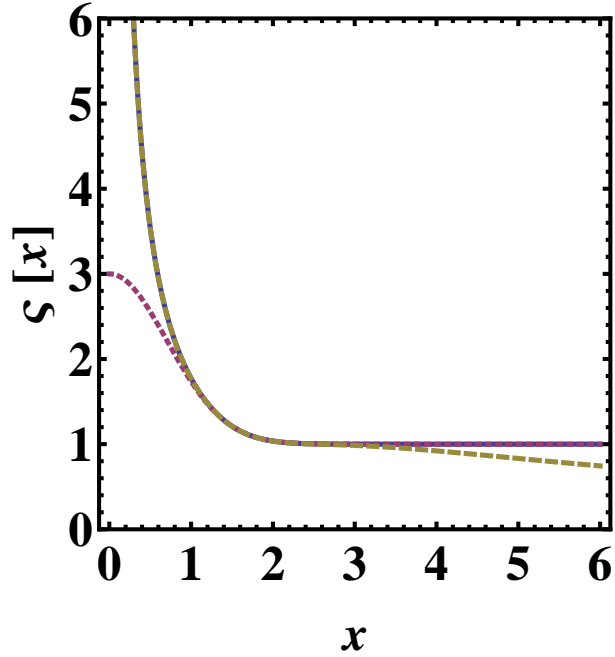}
\end{center}
\caption{Asymptotic values of the function $\Sf(x)$. The plot shows that $\Sf(x)$ can be well approximated at low temperatures ($x<\sqrt{\pi}$) and high temperatures ($x>\sqrt{\pi}$) by the approximate expressions.\label{Sfunc}}
\end{figure}
The pressure at this order is therefore given by
\be
P_1 = -3\lambda \left( \frac{M^6 T^2}{2^6 \pi^3} \right)
\Sf^2\left(\frac{2\pi T}{M}\right)\,.
\label{2loop-pressure}
\ee

Remarkably, one can write $\Sf(x)$  as: \be \label{Sf-duality}
\Sf(x) = \sum_{n=-\infty}^{\infty} {\rm e}^{-n^2x^2}=
\frac{\sqrt{\pi}}{x} \sum_{m=-\infty}^{\infty} {\rm
e}^{-\frac{m^2\pi^2}{x^2}}= \frac{\sqrt{\pi}}{x}\,\Sf\LF{\pi\over
x}\RF \,. \ee The first equality shows explicitly the contribution
of the $n^{\rm{th}}$ thermal mode. In contrast to a standard quantum
field theory, one can see that the higher thermal modes are strongly
suppressed at high temperatures.  When $x\rightarrow\infty$, the
leading term is given by the zero mode ($n=0$). In this limit, the
next contribution is given by the first modes ($n=+1$ and $n=-1$).
Interestingly, Eq. (\ref{Sf-duality}) suggests that at low
temperatures the partition function can be interpreted as arising
from a {\em different} set of modes.  These modes are not
proportional to the temperature, but to the inverse of the
temperature. When $x\rightarrow 0$, the leading term is given by the
inverse zero mode ($m=0$). The next to leading order contribution is
given by the first inverse modes ($m=+1$ and $m=-1$), and so on.

One can use the property given by Eq. (\ref{Sf-duality}) to show that
\be
Z_1(T)=Z_1\left(\frac{{T_c}^2}{T}\right) \,,
\label{duality}
\ee
with $T_c \equiv M/2\sqrt{\pi}$.
The existence of such a duality has indeed been verified in several stringy computations, such as \cite{atick,Polchinski:1985zf,bala,dienes}.

\begin{figure}[bt]
\begin{center}
\resizebox{7.0cm}{!} {\includegraphics{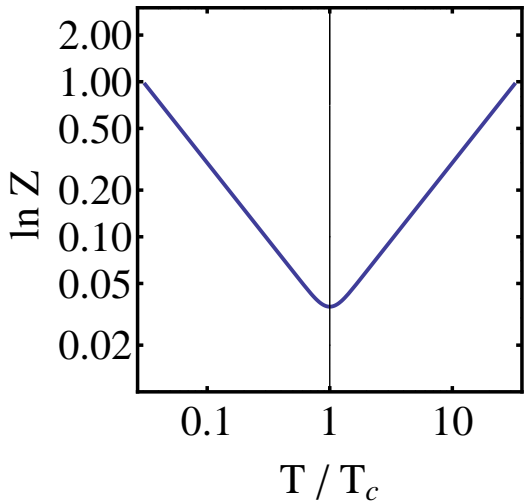}}
\caption{Partition function of the p-adic string model for $p=3$ in
4 space-time dimensions (arbitrary units).  The symmetry of $Z_1(T)$
with respect to $T_c$ shows the exact realization of the thermal
duality at this order in $\lambda$.}
\end{center}
\label{Zofunc}
\end{figure}
Since $\Sf(x)$ can be represented as a series of exponentials, one can obtain excellent high and low temperature approximations to $\Sf(x)$.
The limiting forms of this function are
\be
\Sf(x) \rightarrow \frac{\sqrt{\pi}}{x}
\left[1+2\,{\rm e}^{-\pi^2/x^2}\right] \, ,
 \;\;\; x \ll 1\,,
\label{Tlow}
\ee
and
\be
\Sf(x) \rightarrow 1 + 2\,{\rm e}^{-x^2} \, ,
 \;\;\; x \gg 1\,,
\label{Thigh}
\ee
corresponding to the low and high temperature limits, respectively (see Figure \ref{Sfunc}).  In particular, one finds that
\be
\ln Z_1 \ra \left\{\begin{array}{llc} &=-\La V/T\,,& T\ll M \\
&=-4\pi\La T V/ M^2\,,& T\gg M
\end{array}\Rd
\ee
where $\La\equiv 3\lambda T_c^8$ is the cosmological constant.  This is precisely the kind of asymptotic behavior that has been suggested in the stringy literature \cite{atick,dienes}.

In \cite{prl-old} we discussed the thermodynamic implications of
this result (\ref{2loop-pressure}). We found that at low
temperatures, in addition to a negative cosmological constant, the
thermal properties of the $p$-adic fluid resembles that of a
pressureless dust. In this low density phase the temperature changes
very slowly (logarithmically) with the change of energy density.
Qualitatively, this is also what has been argued from the string
theory side \cite{jain,vafa,robert}; in the Hagedorn phase the
energy density is dominated by the most massive string states and
hence behaves as a pressureless fluid, with the temperature
remaining almost a constant.  However, unlike \cite{jain}, our
partition function does not suffer from the problem of having a
negative heat capacity, and gives way to a $P/\rho = \oa\approx 1$
stiff fluid phase at high temperature.

In this context, we note that if $g_o^2\ll 1$, we can describe the
physics near the Hagedorn temperature $T_c$ by using the partition
function at first order in $\lambda$. We will see this explicitly
when we compare (\ref{Z2loop}) with the order $\lambda^2$
contributions in section \ref{sec:higher-loops}. From Eq.
(\ref{Z2loop}) and Figure \ref{Zofunc}, we conclude that this is a
smooth transition and not a first order phase transition, as has
been conjectured from string theory~\cite{atick}. More recently, the
possibility of such a smooth transition  has also been
suggested~\cite{chaudhuri,dienes}, although in \cite{dienes} a
different higher temperature was identified to correspond to  much
milder phase transitions in some  supersymmetric string models.
Finally, we point out that to this order, since $\lambda$ is defined
negative for the p-adic model, the entropy density is positive, and
that makes it consistent. On the other hand, the vacuum energy
density or cosmological constant is negative, which is not what is
observed in nature.

\section{Self-Energy, Ghosts and a Positive Cosmological Constant}
\label{sec:self-energy}

Higher derivative theories are typically
plagued with ghosts. For instance, a fourth order theory has a
propagator of the form $[p^4+Bp^2+C]^{-1}$ where $B$ and $C$ are
some constants. In general, such a propagator has two poles
corresponding to two different physical degrees of
freedom\footnote{There can be a double pole, but such theories are
also not consistent~\cite{smilga}.}.  It is easy to check that at
least one of them is ghost-like; it has the wrong sign for the
residue at the pole.  One of the virtues of $p$-adic type models is
that the modified higher derivative propagator has no poles, \ie
there are no perturbative states, ghosts or otherwise. However, this
statement is true only at the tree-level. Do quantum corrections
spoil this property?

The lowest order non-zero contribution to the partition function
gives rise to an order $\lambda$ contribution to the self-energy
$\Pi$, defined by \be D^{-1} = D_0^{-1} + \Pi \, , \ee where $D$ is
the full propagator.  The standard computation gives \ba
\Pi_1 &=& 12 \lambda T \sum_n \int \frac{d^3k}{(2\pi)^3} D_0(\omega_n,{\bf k}) \nonumber \\
&=& 12 \lambda T
\left(\frac{M}{2\sqrt{\pi}}\right)^3 \Sf\left( \frac{2\pi T}{M}\right)\,.
\ea
This is dimensionless, just as the propagator is. However, we note the reappearance of a pole: apparently the ghost comes back to haunt us! In~\cite{minahan} a similar result was also obtained. One possible interpretation of the pole is in terms of massive closed string states~\cite{minahan}, but here we adopt a different point of view. We realize that the appearance of a pole in the propagator in the complex $p_0$ plane at $T=0$ can be avoided by adding a counter-term to the Lagrangian of the form $-\thalf \gamma \phi^2$ and adjusting $\gamma$ to cancel the self-energy contribution.  This leads to
\be
\gamma = - \frac{3 \lambda M^4}{4\pi^2}\,.
\ee
The coefficient $\gamma$ must be adjusted order by order to cancel the self-energy contributions. Thus, our prescription provides a unique way of extending the consistency of the tree-level $p$-adic theory to all orders in quantum loops.

In addressing the problem of ghosts, we discover a rather remarkable consequence: the cosmological constant becomes {\em positive}.  This is because the counter-term also contributes to the pressure at order $\lambda$.  The contribution is
\be
-\thalf \gamma T \sum_n \int \frac{d^3k}{(2\pi)^3} D_0(\omega_n,{\bf k})
= \frac{3 \lambda M^4}{8\pi^2} T
\left(\frac{M}{2\sqrt{\pi}}\right)^3 \Sf\left( \frac{2\pi T}{M}\right)
\, ,
\ee
which results in a total pressure of
\be
P_1 = -3 \lambda \left( \frac{M^2}{4\pi} \right)^4 \frac{2\sqrt{\pi}T}{M}
\Sf\left( \frac{2\pi T}{M}\right) \left[\frac{2\sqrt{\pi}T}{M}
\Sf\left( \frac{2\pi T}{M}\right) - 2 \right] \, .
\ee
This is written is such a way as to make clear both the zero and high temperature limits.  A negative value of $\lambda$ leads to a positive vacuum energy or cosmological constant.
\be
\La=- 3 \lambda\left( \frac{M^2}{4\pi} \right)^4
\ee
In addition, the entropy density is a monotonically increasing function of temperature, making the approximation in this theory thermodynamically
self-consistent.

\section{Higher Order Diagrams and Limits to Perturbation Theory}
\label{sec:higher-loops}

In this section we explore contributions to higher order in $\lambda$ than previously considered.  First we compute the diagrams which are of order $\lambda^2$.  Then we sum the infinite set of ring diagrams and the infinite set of necklace diagrams.

\subsection{Free energy at second order}

\begin{figure}
\vspace*{0cm}
\centerline{\mbox{\epsfxsize=5.0 cm\epsfbox{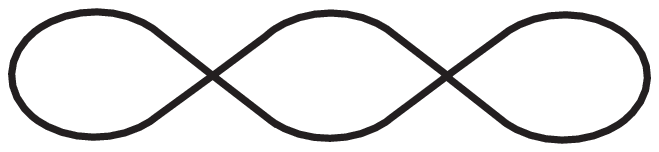}}
\hspace*{2.0 cm}
\mbox{\epsfxsize=2.5 cm\epsfbox{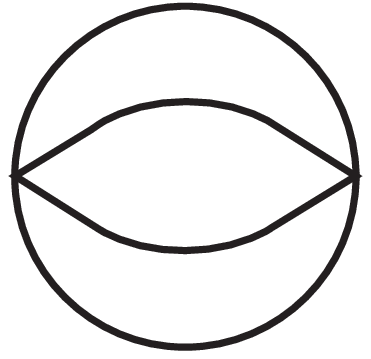}}}
\caption{\footnotesize{Diagrams that contribute at second order in $\lambda$. The one on the left is a necklace diagram, the one on the right is a sunset diagram.}}
\label{neckandsun}
\end{figure}

It has been argued that the duality relation (\ref{duality}) must be broken when nonperturbative effects are included \cite{atick,bala}. We find that in our case it is broken at the next order in $\la$.  Let us compute the contribution that is second order in $\lambda$. One such contribution arises from the necklace diagram shown in Figure \ref{neckandsun}.
\be P_{2,\rm necklace} = 36\lambda^2 \left[T
\left(\frac{M}{2\sqrt{\pi}}\right)^3 \Sf\left( \frac{2\pi
T}{M}\right) \right]^2 \left[T \left(\frac{M}{2\sqrt{2\pi}}\right)^3
\Sf\left( \frac{2\sqrt{2}\pi T}{M}\right) \right] \ee One can see
already under what conditions the perturbative expansion would be
meaningful. For $T\gg M$, $\Sf\ra 1$, and therefore \be P_1\sim (M^3
T)(\la M^3T)\mx{ while } P_{2,\rm necklace} \sim (M^3 T)(\la M^3T)^2
\ . \ee We will show later in this section that this argument can be
extended to arbitrary order. Thus, the perturbative expansion is
only valid when $\la M^3 T\ll 1$. At the other temperature extreme,
$T\ll M$, $\Sf\ra M/T$, and now \be P_1\sim M^4(\la M^4)\mx{ while }
P_{2,\rm necklace} \sim M^4(\la M^4)^2 \ . \ee Thus, at low
temperature the perturbative expansion parameter is $\la M^4\sim
g_o^2$. It is clear already that if $g_o^2$ is small enough we can
trust the order $\lambda$ result all the way up to a temperature
over the critical one $T_c$.

Returning to the problem of calculating the partition function to
second order in $\lambda$, we recognize that there is a
topologically different type of diagram at the same order in
$\lambda$.  This diagram contains two vertices, and every leg of one
vertex is connected to a leg of the other vertex.  This is
oftentimes called a sunset diagram.  Apart from a combinatoric
factor, what needs evaluation is \bd \left[ T \sum_{n_1} \int
\frac{d^3k_1}{(2\pi)^3} D_0(\omega_{n_1},{\bf k}_1) \right]
\cdot\cdot\cdot \left[ T \sum_{n_{4}} \int \frac{d^3k_{4}}{(2\pi)^3}
D_0(\omega_{n_{4}},{\bf k}_{4}) \right] \ed \bd \times (2\pi)^3
\delta({\bf k}_1 + \cdot\cdot\cdot {\bf k}_{4}) \beta
\delta_{n_1+\cdot\cdot\cdot n_{4},0} \, . \ed The momentum
conserving delta function can be expressed in integral form.  Then
the momentum integrals factorize and the remaining integral can be
evaluated with the result \be P_{\rm sunset} =  \frac{3}{2}\lambda^2
\left( \frac{M}{2\sqrt{\pi}} \right)^9 \Ff(T,M) \, . \ee Here the
combinatoric factor has been included and \be \Ff(T,M) = \left[ T
\sum_{n_1} {\rm e}^{-x^2 n_1^2} \right] \cdot\cdot\cdot \left[ T
\sum_{n_{4}} {\rm e}^{-x^2 n_{4}^2} \right] \beta
\delta_{n_1+\cdot\cdot\cdot n_{4},0} \, . \ee It does not seem
possible to obtain the function $\Ff$ in closed form.  However, it
can be rewritten with the aid of the theta function of the third
kind. \be \theta_3(u,q)=\sum_{n=-\infty}^{\infty} q^{n^2} {\rm
e}^{2uni} \, . \ee The sum $\Sf(x)$ encountered so often is just
$\Sf(x) = \theta_3(0,{\rm e}^{-x^2})$.  The theta function can also
be written as an infinite product. \be
\theta_3(u,q)=\prod_{n=1}^{\infty} \left[ 1 + q^{2n-1} \cos(2u) +
q^{2(2n-1)} \right] \left( 1 - q^{2n} \right) \, . \ee The
usefulness arises when we represent the frequency conserving delta
function as an integral. \be \delta_{n_1+\cdot\cdot\cdot n_4,0} =
\int_{-\pi}^{\pi} \frac{d\phi}{2\pi} {\rm e}^{i(n_1 +
\cdot\cdot\cdot n_4)\phi} \, . \ee Then we can write \be \Ff(T,M) =
\int_{-\pi}^{\pi} \frac{d\phi}{2\pi} \left[ \theta_3\left(\thalf
\phi,{\rm e}^{-x^2} \right) \right]^4 \, . \ee We see that the
thermal duality relation Eq. (\ref{duality}) is only verified at the
leading order in $\lambda$, but is broken at order $\lambda^2$.
However, the fact that we can write all the results in terms of
$\theta_3(u,{\rm e}^{-x^2})$, which obeys
\be \theta_3(u,{\rm e}^{-x^2})= \frac{\sqrt{\pi}}{x} {\rm
e}^{-u^2/x^2} \theta_3\left(\frac{i\,\pi \,u}{x^2},{\rm
e}^{-\pi^2/x^2}\right)\,,
\ee
allows an alternative interpretation in terms of inverse modes, but they need to be weighted in a different way.

The low and high temperature limits of the sunset contribution are readily obtained.  The high temperature limit is dominated by all $n_i=0$.  Hence
\be
\Ff(T \gg M,M) = T^3 \, .
\ee
In the low temperature limit the sums can be replaced by integrals and the limits on $\phi$ extended to $\pm \infty$.
\be
\Ff(T \ll M,M) = \frac{1}{2} \left(\frac{M}{2\sqrt{\pi}}\right)^3 \, .
\ee
Therefore
\be
P_{\rm sunset}(T \gg M) = \frac{3}{2} \lambda^2
\left(\frac{M}{2\sqrt{\pi}}\right)^9 T^3 \, .
\ee
and
\be
P_{\rm sunset}(T \ll M) = \frac{3}{4} \lambda^2
\left(\frac{M^2}{4\pi}\right)^6 \, .
\ee

Note that both the necklace and the sunset contribution grows at high temperature as $T^3$ compared to the first order in $\lambda$ contribution which grows as $T^2$.  The ratio is
\be
P_{\rm 2,necklace}/P_1 \sim P_{\rm sunset}/P_1 \sim\lambda M^3 T \sim g_o^2 T/m_s \, .
\ee
This ratio is small and only approaches one when the temperature reaches $T \sim m_s/g_o^2$.

\subsection{Ring diagrams}

In ordinary field theories with massless particles one generally finds a certain class of diagrams that are infrared divergent, and that the divergence becomes more severe with increasing number of loops.  These are the ring diagrams.  Their contribution to the pressure is
\ba
P_{\rm ring} &=& \thalf T \sum_n \int \frac{d^3k}{(2\pi)^3} \sum_{l=2}^{\infty} \frac{1}{l} \left[ -\Pi_1 D_0(\omega_n,{\bf k}) \right]^l
\nonumber \\
&=& - \thalf T \sum_n \int \frac{d^3k}{(2\pi)^3}
\left[ \ln \left( 1 + \Pi_1 D_0 \right) - \Pi_1 D_0 \right] \, .
\ea
In ordinary massless $\phi^4$ theory the propogator is
$D_0 = 1/(\omega_n^2 + {\bf k}^2)$ and $\Pi_1$ is frequency and momentum independent.  The sum of these ring diagrams is proportional to $\lambda^{3/2}$, nonanalytic in the coupling constant.  This nonanalytic contribution arises  from the $n=0$ term in the Matsubara summation.  In the present non-local theory the individual diagrams are convergent and there is no need to sum the series.  If one chooses to sum the series anyway there is a limit to its convergence.  The largest value of $D_0$ is 1.  For the series to converge to the logarithm one therefore needs $-1 < \Pi_1 \le 1$.  For given $\lambda$ and $M$ there is a maximum $T$ for which the series converges to the logarithm.

In fact, it is possible to show that there is no divergence problem from the set of ring diagrams.  Each diagram in the series is convergent. We can evaluate them by using Eq. (\ref{useful1}). In the limit in which $l \rightarrow \infty$, they go as $1/l^{3/2}$.  Hence at large $l$, the terms go as $1/l^{5/2}$.  This series converges much more rapidly than does a logarithm.

\subsection{Necklace diagrams}

There is another set of diagrams that are easily summed which are similar to the set of ladder diagrams in scattering theory.  Start with the Figure \ref{neckandsun} diagram, and keep adding loops to make a string.
\begin{figure}
\begin{center}
\includegraphics[width=3.5in,angle=0]{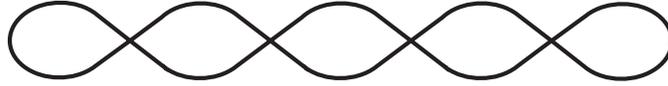}
\caption{A necklace diagram which is fourth order in $\lambda$ for the case $p=3$.\label{necklacep2}}
\end{center}
\end{figure}
The first and last loops are not connected to each other.  This looks like an open necklace for $p=3$, see Figure \ref{necklacep2}.  Including the appropriate combinatoric factors, we find:
\bd
P_{\rm necklace} =
3 \sum_{l=0}^{\infty} (-\lambda)^{l+1} 12^l
\left[T \sum_n \int \frac{d^3k}{(2\pi)^3} D_0(\omega_n,{\bf k}) \right]^2
\ed
\bd
\times
\left[T \sum_n \int \frac{d^3k}{(2\pi)^3} D_0^2(\omega_n,{\bf k})  \right]^l
\ed
\bd
= -3\lambda \left[T \left(\frac{M}{2\sqrt{\pi}}\right)^3
\Sf\left( \frac{2\pi T}{M}\right) \right]^2
\sum_{l=0}^{\infty} \left[ -12\lambda
T \left(\frac{M}{2\sqrt{2\pi}}\right)^3
\Sf\left( \frac{2\sqrt{2}\pi T}{M}\right) \right]^l
\ed
\be
= \frac{-3\lambda \left[T \left(\frac{M}{2\sqrt{\pi}}\right)^3
\Sf\left( \frac{2\pi T}{M}\right) \right]^2}
{1+12 \lambda T \left(\frac{M}{2\sqrt{2\pi}}\right)^3
\Sf\left( \frac{2\sqrt{2}\pi T}{M}\right)} \,.
\ee
Notice that the numerator is just the lowest order result but without the mass counter-term.  To include the latter one just makes a self-energy insertion on either end of the string.  (Making an insertion on both ends is one higher order in $\lambda$ and is just one contribution to the set of super ring diagrams.)  The result is
\be
P_{\rm necklace} = \frac{ -3 \lambda
\left( \frac{M^2}{4\pi} \right)^4 \frac{2\sqrt{\pi}T}{M}
\Sf\left( \frac{2\pi T}{M}\right)
\left[\frac{2\sqrt{\pi}T}{M} \Sf\left( \frac{2\pi T}{M}\right) - 2 \right]}
{1+12 \lambda T \left(\frac{M}{2\sqrt{2\pi}}\right)^3
\Sf\left( \frac{2\sqrt{2}\pi T}{M}\right)} \,.
\label{funone}
\ee

This expression has some very interesting properties when $\lambda < 0$.  As before, the vacuum energy density is positive.  The entropy density is positive at all temperatures.  In addition, there is a maximum temperature determined by the vanishing of the denominator.  We can try to interpret
this fact as arising due to  the potential being unbounded from below for large values of $\phi$. Indeed, the potential is
\be
U(\phi) = \thalf \phi^2 + \lambda \phi^4 \,.
\ee
ignoring the correction from $\gamma \phi^2$.  The maximum height of the potential is
\be
U_{\rm max} = - \frac{1}{16 \lambda} \,.
\ee
For the sake of estimation, use the lowest order expression for the energy density in the limit $T > M$.  It is
\be
\epsilon_1 = -3\lambda \left( \frac{M^2}{4\pi} \right)^3 T^2 \,.
\ee
When the temperature is high enough the states will spill over the top of the potential.  One can estimate this temperature by equating $U_{\rm max}$ and $\epsilon_1$.  This gives a so-called critical temperature of
\be
T^2 = \frac{1}{6\lambda^2} \left( \frac{2\pi}{M^2} \right)^3 \,.
\ee
On the other hand we can find the temperature for which the denominator of Eq. (\ref{funone}) vanishes.  In the same high $T$ approximation we get
\be
T^2 = \frac{4}{9\lambda^2} \left( \frac{2\pi}{M^2} \right)^3 \,.
\ee
Parametrically these two estimates are identical and equivalent to a limiting temperature of order $T \sim m_s/g_o^2$.  At this temperature there is no longer an identifiable expansion parameter. If one wants to consider temperatures higher than the limiting temperature one finds a negative entropy density indicating that the system is unstable ($\phi=0$ no longer being a stable vacuum).

Our loop calculations suggest some general power counting arguments.
For low temperature an $l$-loop graph is suppressed as
$g_o^{2(l-1)}$, while at high temperature the expansion parameter is
$  (g_o^2\,T/m_s)^{l-1}$. What is left undetermined of course is the
coefficient in front of the power, and one would need to sum all the
diagrams at a given loop level to determine this. For the necklace
diagrams this coefficient remained a constant with increasing loops
leading to the singular expression for the pressure (\ref{funone}).
However, it is possible that when one accounts for all the diagrams,
the coefficients may decrease faster than $1/l$ with increasing
loops and therefore lead to no divergences. However, in the light of
the physical argument we have presented which relates the existence
of the tachyon with the occurrence of the singular behavior, this
seems rather unlikely. In a realistic string model one does not
expect real tachyons to be present. Accordingly, we expect that
either the presence of supersymmetry and/or additional interactions
will ameliorate the above singularity.  In fact, soliton
contributions become important at the same temperature
\cite{Biswas:2010yx}.

\section{Arbitrary Even Powered Potentials}
\label{sec:higher-p}

The above analysis can easily be extended to an interaction term of the form
$-\lambda \phi^{2N}$ where $N = (p+1)/2$.  (The previous analysis corresponds to $N=2$.)  Note that $\lambda$ has energy dimension $-4(N-1)$.

\subsection{Lowest order diagrams}

\begin{figure}
\begin{center}
\includegraphics[width=2.0in,angle=0]{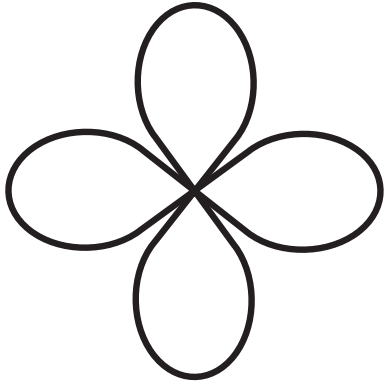}
\caption{This four loop diagram is the leading order in $\lambda$ for the case $p=7$.\label{4loopsp4}}
\end{center}
\end{figure}

Consider the partition function to first order in $\lambda$.  It involves a diagram with one vertex and $2N$ legs which can be connected into $N$ loops.  See Figure \ref{4loopsp4}.
Taking into account the combinatoric factor gives \be \ln Z_1 =
(2N-1)!!(-\lambda) \beta V \left[ T \sum_n \int
\frac{d^3k}{(2\pi)^3} D_0(\omega_n,{\bf k}) \right]^N \ee Similarly,
the first order contribution to the self-energy is \be \Pi_1 =
2N(2N-1)!! \lambda \left[ T \sum_n \int \frac{d^3k}{(2\pi)^3}
D_0(\omega_n,{\bf k}) \right]^{N-1} \ee We add the counter-term
$-\thalf \gamma \phi^2$ and adjust the coefficient to cancel the
self-energy contribution at $T=0$:
\be \gamma = -2N(2N-1)!! \lambda \left( \frac{M^2}{4\pi}
\right)^{2(N-1)} \, .
\ee
This leads to the total pressure at first order
\be P_1 = -(2N-1)!! \lambda \left( \frac{M^2}{4\pi} \right)^{2N}
\frac{2\sqrt{\pi}T}{M} \Sf\left( \frac{2\pi T}{M}\right) \left[
\left(\frac{2\sqrt{\pi}T}{M}  \Sf\left( \frac{2\pi
T}{M}\right)\right)^{N-1} - N \right] \, . \ee
The vacuum energy density is
\be \epsilon_{\rm vac} = -(N-1)(2N-1)!! \lambda \left(
\frac{M^2}{4\pi} \right)^{2N} \, ,
\label{pvacuum}
\ee
and the high temperature limit of the pressure is
\be
P_1 = -(2N-1)!! \lambda \left( \frac{M}{2\sqrt{\pi}}
\right)^{3N} T^N  \, ,
\ee
which is notable for its proportionality to $T^N$.

\subsection{Necklace diagrams}

\begin{figure}
\begin{center}
\includegraphics[width=4.0in,angle=0]{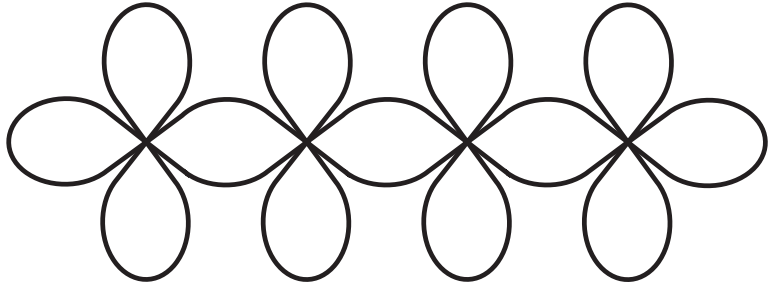}
\caption{The necklace diagram of fourth order in $\lambda$ for the case $p=7$.
\label{necklacep4}}
\end{center}
\end{figure}

The necklace diagrams are obtained by connecting each vertex with
two legs. When $N=2$ the end vertices have one closed loop attached
to them while the interior vertices have none.  When $N>2$ the end
vertices have $N-1$ closed loops attached to them while the interior
vertices have $N-2$ closed loops attached (see Figure
\ref{necklacep4}).
The pressure contribution of the set of such diagrams is
\bd P_{\rm necklace} = (-\lambda) (2N-1)!! \left[T \sum_n \int
\frac{d^3k}{(2\pi)^3} D_0(\omega_n,{\bf k}) \right]^N \times \ed \bd
\sum_{l=0}^{\infty} \left\{ (-\lambda) 4(2N-1)!! \left[T \sum_n \int
\frac{d^3k}{(2\pi)^3} D_0(\omega_n,{\bf k}) \right]^{N-2} T \sum_n
\int \frac{d^3k}{(2\pi)^3} D_0^2(\omega_n,{\bf k}) \right\}^l \ed
\be = \frac{- (2N-1)!! \lambda \left[T
\left(\frac{M}{2\sqrt{\pi}}\right)^3 \Sf\left( \frac{2\pi
T}{M}\right) \right]^N} {1+ 4(2N-1)!! \lambda T
\left(\frac{M}{2\sqrt{2\pi}}\right)^3 \Sf\left( \frac{2\sqrt{2}\pi
T}{M}\right) \left[T \left(\frac{M}{2\sqrt{\pi}}\right)^3 \Sf\left(
\frac{2\pi T}{M}\right) \right]^{N-2}} \ee

After taking into account self-energy corrections on the loops attached to the end vertices we finally get
\be P_{\rm necklace} = \frac{ -(2N-1)!! \lambda \left(
\frac{M^2}{4\pi} \right)^{2N} \frac{2\sqrt{\pi}T}{M} \Sf\left(
\frac{2\pi T}{M}\right) \left[ \left(\frac{2\sqrt{\pi}T}{M}
\Sf\left( \frac{2\pi T}{M}\right)\right)^{N-1} - N \right]} {1+
4(2N-1)!! \lambda T \left(\frac{M}{2\sqrt{2\pi}}\right)^3 \Sf\left(
\frac{2\sqrt{2}\pi T}{M}\right) \left[T
\left(\frac{M}{2\sqrt{\pi}}\right)^3 \Sf\left( \frac{2\pi
T}{M}\right) \right]^{N-2}} \, . \ee

As before, there is a maximum temperature determined by the vanishing of the denominator.  As with the 3-adic case, we can ascribe this singular
behavior to the presence of the tachyon in the potential.  The important point is that the potential is unbounded from below for large values of $\phi$.  The potential is
\be
U(\phi) = \thalf \phi^2 + \lambda \phi^{2N}
\ee
ignoring the correction from $\gamma \phi^2$.  The maximum height of the potential is
\be
U_{\rm max} = \frac{N-1}{2N} \left(\frac{-1}{2N \lambda}\right)^{1/(N-1)}
\ee
For the sake of estimation, use the $N$-loop expression for the energy density in the limit $T > M$.  It is
\be
\epsilon_1 = -\lambda (N-1)(2N-1)!! \left( \frac{M}{2\sqrt{\pi}}
 \right)^{3N} T^N
\ee
When the temperature is high enough the states will spill over the top of the potential.  One can estimate this temperature by equating $U_{\rm max}$ and $\epsilon_1$.  This gives a so-called critical temperature of
\be
T = \left(\frac{1}{(2N-1)!!}\right)^{1/N}
\left( \frac{-1}{2N\lambda} \right)^{1/(N-1)}
\left( \frac{2\sqrt{\pi}}{M} \right)^3
\ee

On the other hand we can find the temperature for which the denominator of the necklace pressure vanishes.  In the same high $T$ approximation we get

\be
T = \left( \frac{-1}{\sqrt{2}(2N-1)!!\lambda} \right)^{1/(N-1)}
\left( \frac{2\sqrt{\pi}}{M} \right)^3
\ee
Parametrically these two estimates are identical.

\subsection{Sunset diagrams}

\begin{figure}
\begin{center}
\includegraphics[width=2.0in,angle=0]{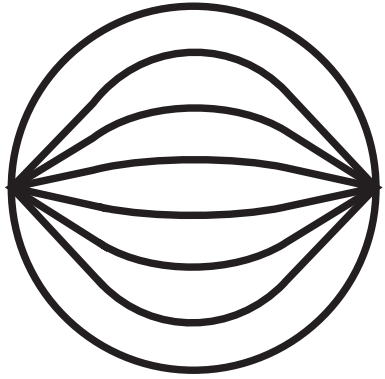}
\caption{This four loop diagram is the sunset diagram for the case $p=7$.
\label{sunsetp4}}
\end{center}
\end{figure}

The sunset diagram has two vertices, and every leg of one vertex is connected to a leg of the other vertex.  This is the generalization of the usual sunset diagram from $\phi^4$ theory.  See Figure \ref{sunsetp4}.
Apart from a combinatoric factor, what needs evaluation is
\bd
\left[ T \sum_{n_1} \int \frac{d^3k_1}{(2\pi)^3}
D_0(\omega_{n_1},{\bf k}_1) \right] \cdot\cdot\cdot
\left[ T \sum_{n_{2N}} \int \frac{d^3k_{2N}}{(2\pi)^3}
D_0(\omega_{n_{2N}},{\bf k}_{2N}) \right]
\ed
\bd
\times (2\pi)^3 \delta({\bf k}_1 + \cdot\cdot\cdot {\bf k}_{2N}) \beta \delta_{n_1+\cdot\cdot\cdot n_{2N},0} \, .
\ed
The momentum conserving delta function can be expressed in integral form.  Then the momentum integrals factorize and the remaining integral evaluated with the result
\be
P_{\rm sunset} = \lambda^2 \frac{(2N)!}{2(2N)^{3/2}}
\left(\frac{M}{2\sqrt{\pi}}\right)^{3(2N-1)} \Ff(T,M) \, .
\ee
Here the combinatoric factor has been included and
\be
\Ff(T,M) = \left[ T \sum_{n_1} {\rm e}^{-x^2 n_1^2} \right] \cdot\cdot\cdot
\left[ T \sum_{n_{2N}} {\rm e}^{-x^2 n_{2N}^2} \right] \beta \delta_{n_1+\cdot\cdot\cdot n_{2N},0} \, .
\ee
Analogous to the case $p=3$ we can write
\be
\Ff(T,M) = \int_{-\pi}^{\pi} \frac{d\phi}{2\pi}
\left[ \theta_3\left(\thalf \phi,{\rm e}^{-x^2} \right) \right]^{2N}
\ee

The low and high temperature limits are readily obtained.  The high temperature limit is dominated by all $n_i=0$.  Hence
\be
\Ff(T \gg M) = T^{2N-1}
\ee
In the low temperature limit the sums can be replaced by integrals and the limits on $\phi$ extended to $\pm \infty$.
\be
\Ff(T \ll M) = \frac{1}{\sqrt{2N}} \left(\frac{M}{2\sqrt{\pi}}\right)^{2N-1}
\ee
Therefore
\be
P_{\rm sunset}(T \gg M) = \lambda^2 \frac{(2N)!}{2(2N)^{3/2}}
\left(\frac{M}{2\sqrt{\pi}}\right)^{3(2N-1)} T^{2N-1}
\ee
and
\be
P_{\rm sunset}(T \ll M) = \lambda^2 \frac{(2N)!}{8N^2}
\left(\frac{M^2}{4\pi}\right)^{2(2N-1)} \, .
\ee
These are precisely the same asymptotic behaviors that one obtains from the 3-loop necklace diagram.

In particular, both the sunset and necklace contributions grow at high temperature as $T^{2N-1}$ compared to the lowest order in $\lambda$ contribution which grows as $T^N$.  The ratio is
\be
P_{\rm 2,necklace}/P_1 \sim P_{\rm sunset}/P_1 \sim \lambda \left( M^3 T \right)^{N-1} \, .
\ee
This ratio is small and only approaches one when the limiting temperature, as defined by the vanishing of the denominator in the formula for the necklace contribution, is reached.  Hence these 3-loop contributions can be neglected since it is important only at unreachable temperatures where other contributions undoubtedly become important too.  Also notice that the two contributions do not cancel each other because they both have positive coefficients.

\section{$D$ dimensional $p$-adic Theory and the Vacuum Energy}
\label{sec:cc}

In this section our aim is to calculate the
cosmological constant that one obtains in $p$-adic string theories
for any general value of $p$, and in particular to see how it
depends on the different scales and couplings.  First of all, we
note that $p$-adic strings can be formulated in arbitrary space time
dimension; however, what we observe today is the effective four
dimensional cosmological constant. Thus we proceed in two steps. In
the first step, we compute the $D$ dimensional cosmological constant
for arbitrary $p$. For this purpose we are going to assume that the
extra $\Dn=D-4$ dimensions are all compactified on a circle of
radius $R$. In the second step, we obtain the dimensionally reduced
cosmological constants. In particular, the
results of the computation simplifies considerably in two opposite
limits: (a) $RM\ll 1$, which corresponds to the usual dimensional
reduction where one only includes the zero-modes, and (b) $RM\gg 1$,
the t-dual limit which is interesting for TeV scale
phenomenology. We therefore can obtain estimates of the
cosmological constant in terms of the physical parameters in these
two limits.

\subsection{3-adic in arbitrary dimensions}

\subsubsection{Higher dimensional vacuum energy}

We will first illustrate our method for 3-adic theory and later
generalize to the $p$-adic case. Starting from an arbitrary $D$
dimensional theory we are led to the action
\be
\label{generals} S =
\int_0^{\beta}d\tau\int_0^{2\pi R} d^{\Dn}y \int d^3x \left[ -\thalf
\phi {\rm e}^{-(\partial^2/\partial \tau^2 + \nabla^2_y+
\nabla^2_x)/M^2} \phi - \lambda \phi^4 \right] \, , \ee where here
$\phi = \phi({\bf x},{\bf y},\tau)$.  Notice that $\phi$ now has
mass dimension $D/2$, while \be \la=-{m_s^{-D}g_0^2\over 18} \ee has
mass dimension $-D$. All the calculations that were done for four
dimensions now go through except that instead of one compact
dimension (temperature) we now have $\Dn+1$ of them. Accordingly, we
have to replace
\be
\bt \ra \bt (2\pi R)^{\Dn} \ee and \ba \sum_n
\int \frac{d^3k}{(2\pi)^3} {\rm e}^{-(\omega_n^2 + {\bf k}^2)/M^2}
&\ra& \sum_n \sum_{\{n_i\}}\int \frac{d^3k}{(2\pi)^3}
{\rm e}^{-(\omega_n^2 + \sum_i \omega_{n_i}^2+{\bf k}^2)/M^2} \nonumber \\
&\equiv& \sum_n \sum_{\{n_i\}}\int \frac{d^3k}{(2\pi)^3}
D_{\Dn}(\{\oa_n\},{\bf k})
\ea
where
\be
\oa_{n_i}={n_i\over R} \,.
\ee
According to the generalization, the $D$ dimension partition function at order $\lambda$ reads
\be
\ln Z_1 = 3(-\lambda) \beta V (2\pi R)^{\Dn} \left[ T (2\pi R)^{-\Dn}\sum_n  \int \frac{d^3k}{(2\pi)^3}
D_d(\omega_n,{\bf k}) \right]^2\,.
\ee
It is useful to define
\be
\sum_n \int \frac{d^3k}{(2\pi)^3} D_d(\omega_n,{\bf k}) =
\left(\frac{M}{2\sqrt{\pi}}\right)^3 \Sf_d\left( \frac{2\pi T}{M},\frac{1}{RM}\right)\,,
\label{useful2}
\ee
analogous to (\ref{useful1}) valid for four dimensions. It is easy to check that
\be
\Sf_d(x,y) = \Sf(x)\Sf^d(y)\ ,
\ee

One can now follow all the steps leading to the positive cosmological constant. The pressure in $D$ dimensions, coming from the partition function  is
\be
P_1 = 3(-\lambda)\left(\frac{M}{2\sqrt{\pi}}\right)^6  \left[ T (2\pi R)^{-\Dn} \Sf_d\left( \frac{2\pi T}{M},\frac{1}{RM}\right)\right]^2\,.
\ee
while the self-energy is given by
\ba
\Pi_1 &=& 12 \lambda T(2\pi R)^{-\Dn} \sum_n \int \frac{d^3k}{(2\pi)^3} D_d(\{\omega_n\},{\bf k}) \nonumber \\
&=& 12 \lambda T(2\pi R)^{-\Dn}
\left(\frac{M}{2\sqrt{\pi}}\right)^3 \Sf_d\left( \frac{2\pi T}{M},\frac{1}{RM}\right)\,.
\ea
Again, the appearance of a pole can be avoided by adding a counter-term to the Lagrangian of the form $-\thalf \gamma \phi^2$ and adjusting $\gamma$ to cancel the self-energy contribution.
This leads to
\be
\gamma = - \frac{3 \lambda M^4}{4\pi^2}(2\pi R)^{-\Dn}\Sf^d\left(\frac{1}{RM}\right)\,.
\ee
The contribution of the counter-term to the pressure is given by
\ba
P_c &=& -\thalf {\gamma T\over (2\pi R)^{\Dn}}\sum_n \int \frac{d^3k}{(2\pi)^3} D_d(\{\omega_n\},{\bf k}) \nonumber \\
&=& \frac{3 \lambda M^4 T}{8\pi^2(2\pi R)^{\Dn}}
\left(\frac{M}{2\sqrt{\pi}}\right)^3 \Sf_d\left( \frac{2\pi T}{M},\frac{1}{RM}\right)\,.
\ea
The total two loop pressure then reads
\be
P_1 = -3 \lambda \left( \frac{M^2}{4\pi} \right)^4 \frac{2\sqrt{\pi}T}{M}
\Sf\left( \frac{2\pi T}{M}\right){\Sf^{2d}\left(\frac{1}{RM}\right)\over (2\pi R)^{2\Dn}}
\left[\frac{2\sqrt{\pi}T}{M} \Sf\left( \frac{2\pi T}{M}\right) - 2 \right]\,.
\ee
The $D$ dimensional cosmological constant is thus given by
\be
\La_D=- 3 \lambda\left( \frac{M^2}{4\pi} \right)^4{\Sf^{2d}\left(\frac{1}{RM}\right)\over (2\pi R)^{2\Dn}} \, .
\ee
\subsubsection{Dimensional reduction}

We start by noting that $D$ and four dimensional pressures and energy densities are just related via
\ba
P_{4\; {\rm dimensions}} &=& (2\pi R)^{\Dn}P_{D\; {\rm dimensions}} \nonumber \\
\La_{4\; {\rm dimensions}} &=& (2\pi R)^{\Dn} \La_{D\; {\rm dimensions}}  \, .
\ea
We conclude that
\be
\La=- {3 \lambda\Sf^{2d}\left(\frac{1}{RM}\right)\over (2\pi R)^{\Dn}}\left( \frac{M^2}{4\pi} \right)^4
\ee

It is interesting to look at the two opposite limits. For $MR\ll 1$, $\Sf(1/RM)$ is approximately 1, and we have
\be
\La=- {3\la\over (2\pi R)^{\Dn}}\left( \frac{M^2}{4\pi} \right)^4 = {g_o^2\over 6m_s^4(2\pi R)^{\Dn}}\left( \frac{M^2}{4\pi} \right)^4 \for R\ll M^{-1}
\ee
We would have got exactly the same expression by first performing a dimensional reduction of the $D$ dimensional 3-adic action, and then realizing that the $\la$ in (\ref{3adic}).  has to be rescaled by
$$\la\ra {\la\over (2\pi R)^{\Dn}}$$
On the other hand when $MR\gg 1$, using the approximation $\Sf=(\sqrt{\pi}MR)^d$ we have
\be
\La=- {3 \lambda M^{2D}\over(4\pi)^4} \LF{R\over 2}\RF^{\Dn}={g_o^2M^{2D}\over 6m_s^{10} (4\pi)^4} \LF{R\over 2}\RF^{\Dn}\for R\gg M^{-1}
\ee

Physically it is more clarifying to re-express the cosmological
constant in terms of the Planck and string scales. To do this we
need to compactify the $D$ dimensional gravitational action. Again
considering $\Dn$ compact toroidal extra dimensions, it is  possible
to perform a straight forward dimensional reduction of the 10
dimensional supergravity action. \be S_{{\rm sugra}}= \frac{1}{16\pi
G_{10}} \int d^{10}x \, \sqrt{-g} \, {\rm e}^{-2\Phi}\cR_{10}+
\cdot\cdot\cdot \, . \ee Here $\Phi$ is the dilaton which we assume
to be stabilized and as usual identified with the string coupling
constant, $g_o={\rm e}^{\Phi}$, and $G_{10} = (2\pi^2)^3
\alpha^{\prime 4}$ is the 10 dimensional gravitational constant.
Performing the standard dimensional reduction we find \be S_{{\rm
sugra}} = \frac{V_6}{16\pi (2\pi^2)^3 g_o^2 \alpha^{\prime 4}} \int
d^4x \, \sqrt{-g} \, \cR_4 = \frac{1}{16\pi G_N} \int d^4x \,
\sqrt{-g} \, \cR_4 \ee where $G_N$ is Newton's constant and $V_6$ is
the volume of the compactified space, in this case $(2\pi R)^6$.
Thus one finds the well known relation between the 10 dimensional
string scale and the Planck mass. \be M_{\rm P}^2 = \frac{1}{G_{\rm
N}} = \frac{V_6}{(2\pi^2)^3 g_o^2 \alpha^{\prime 4}} = \frac{2V_6
m_s^8}{\pi^6 g_o^2} \ee For the two limits the cosmological constant
reads \be \La \sim \left\{\begin{array}{lll}
 (m_s/M_P)^6 M_P^4 &\for& R\ll M^{-1}\\
 (m_sR)^{12} (m_s/M_P)^6 M_P^4 &\for& R\gg M^{-1} \end{array}
\Rd
\ee
An intriguing feature of the above expressions is that the cosmological constant is suppressed by some power of the ratio of the string scale to the Planck scale. In particular there is considerable phenomenological interest when $m_s \sim$ TeV, but we see by power counting that for the 3-adic case, the suppression is not sufficient.

\subsection{Arbitrary $N$}

We are now ready to compute the cosmological constant for arbitrary
values of $N$.  The analysis with $N=2$ can easily be extended to an
interaction term of the form
\be -\lambda \phi^{2N}\with \la\equiv
-{m_s^{-D(N-1)}g_p^{2(N-1)}\over 2N} \ee
that arises from the $p$-adic action (\ref{generals}). We will focus
only on the case when $p$ is an odd positive integer.  We note that
$\lambda$ has energy dimension $-D(N-1)$.

Consider the partition function to first order in $\lambda$.  It involves a diagram with one vertex and $2N$ legs which can be connected into $N$ loops.  Taking into account the combinatoric factor gives
\ba
\ln Z_1 &=& (2N-1)!!(-\lambda) \beta(2\pi R)^d V \left[ T (2\pi R)^{-d}\sum_{\{n_i\}} \int \frac{d^3k}{(2\pi)^3}
D_d(\{\omega_{n_i}\},{\bf k}) \right]^N\non\\
&=& (2N-1)!!(-\lambda) \beta(2\pi R)^d V \left[
\left(\frac{M}{2\sqrt{\pi}}\right)^3T \Sf\left(\frac{2 \pi
T}{M}\right)\LF{\Sf(\frac{1}{M\,R})\over 2\pi R}\RF^{d} \right]^N
\ea leading to a $D$ dimensional pressure \be P_1 =
(2N-1)!!(-\lambda) \left[ \left(\frac{M}{2\sqrt{\pi}}\right)^3T
\Sf\left(\frac{2 \pi T}{M}\right)\LF{\Sf(\frac{1}{M\,R})\over 2\pi
R}\RF^{d} \right]^N \ee

As before one can also calculate the first order contribution to the self-energy:
\ba
\Pi_1& =& 2N(2N-1)!! \lambda \left[ T (2\pi R)^{-d}\sum_{\{n_i\}} \int \frac{d^3k}{(2\pi)^3}
D_d(\{\omega_{n_i}\},{\bf k}) \right]^{N-1} \nonumber \\
&=&2N(2N-1)!! \lambda \left[ T\Sf\left(\frac{2 \pi T}{M}\right)
\left(\frac{M}{2\sqrt{\pi}}\right)^3 \LF{\Sf(\frac{1}{M\,R})\over
2\pi R}\RF^{d} \right]^{N-1}
\ea
Again, we can add the counter-term $-\thalf \gamma \phi^2$ and adjust the coefficient to cancel the self-energy contribution at $T=0$.
\be
\gamma = -2N(2N-1)!! \lambda \left[ \LF\frac{M}{2\sqrt{\pi}}\RF^4 \LF{\Sf(\frac{1}{M\,R})\over
2\pi R}\RF^{d}\right]^{N-1} \ee This leads to the total pressure \bd
P_1 = -(2N-1)!! \lambda \left( \frac{M^2}{4\pi} \right)^{2N}
\LF{\Sf(\frac{1}{M\,R})\over 2\pi R}\RF^{dN}\frac{2\sqrt{\pi}T}{M}
\Sf\left( \frac{2\pi T}{M}\right) \ed \be \times \left[
\left(\frac{2\sqrt{\pi}T}{M} \Sf\left( \frac{2\pi
T}{M}\right)\right)^{N-1} - N \right] \,
\ee
at first order. In particular, the  high temperature limit of the pressure is \be
P_1 =
-(2N-1)!! \lambda \left( \frac{M}{2\sqrt{\pi}} \right)^{3N}
\LF{\Sf(\frac{1}{M\,R})\over 2\pi R}\RF^{dN}T^N
\ee
which is notable
for its proportionality to $T^N$.

Finally, by looking at the $T\ra 0$ limit we find that the $D$
dimensional vacuum energy density is
\be \La_{\rm vac} = -(N-1)(2N-1)!! \lambda
\LF{\Sf(\frac{1}{M\,R})\over 2\pi R}\RF^{dN}\left( \frac{M^2}{4\pi}
\right)^{2N} \, . \ee
After dimensional reduction we have
\be \La_{4} = -(N-1)(2N-1)!! \lambda (2\pi
R)^d\LF{\Sf(\frac{1}{M\,R})\over 2\pi R}\RF^{dN}\left(
\frac{M^2}{4\pi} \right)^{2N} \, . \ee
We can again look into the two limiting cases.
\be \La=\left\{\begin{array}{lll}
 -(N-1)(2N-1)!! \lambda
\LF 2\pi R\RF^{-d(N-1)}\left( \frac{M^2}{4\pi} \right)^{2N} \sim
\LF{m_s\over M_p}\RF^{2(N+1)} M_p^4 &\for& R\ll M^{-1}\\
-(N-1)(2N-1)!! \lambda (2\pi R)^d\LF{M\over 2\sqrt{\pi} }\RF^{DN}
\sim g_o^{2N}\LF{m_s\over M_p}\RF^{2} M_p^4 &\for& R\gg M^{-1}
\end{array} \Rd \ee

\subsection{Numerical Results for the Cosmological Constant}

We can estimate what parameter choices would reproduce the observed
cosmological constant $\Lambda = (2.3$ meV)$^4$. We will specialize
to the case when $D=6$ \ie, we have six compact extra dimensions.
The Planck mass is related to Newton's constant, the open string
coupling, the Regge slope, and the volume of compactified space by
\be
M_{\rm P}^2 = \frac{1}{G_{\rm N}} = \frac{V_6}{(2\pi^2)^3 g_o^2
\alpha^{\prime 4}} \, ,
\ee
with $m_s^2 = 1/2\alpha^{\prime}$.
Rewriting in terms of the string scale and the radii of the
compactified dimensions we have
\be
M_{\rm P}^2 = \frac{(2\pi
R)^6m_s^8 16}{(2\pi^2)^3 g_o^2 }=\frac{2^7 R^6m_s^8}{ g_o^2 }
\label{planck}
\ee
This imposes a constraint on the three
independent parameters in the theory, $m_s,g_0$ and $R$. In terms of
the same variables, the coupling constant is given by
\be \lambda = -\frac{g_p^{2(N-1)}}{2Nm_s^{10(N-1)}} \, . \ee
We are now ready to estimate the cosmological constant in $p$-adic
theories. Let us first look at the $MR\ll1$ limit. In this case we
have
\be \La=-\la {(N-1)(2N-1)!!\over (2\pi R)^{6(N-1)}}\LF{M^2\over
4\pi}\RF^{2N}\,. \ee
Curiously, one observes that in both the expression for $M_p$ and
$\La$, the same combination of $g_o/R^3$ appears, and hence one can
eliminate them to obtain the value of the string scale required to
obtain the correct hierarchy between the observed cosmological
constant and the Planck scale:
\be {\La\over M_p^4}=Q_<\LF{m_s\over M_p}\RF^{2(N+1)}\with Q_<\equiv
{(N-1)(2N-1)!!\over 2^{N+2}\pi^{8N-6}N(\ln
(2N-1))^{2N}}\LF{2(N-1)\over (2N-1)^2}\RF^{N-1}\,. \ee
For example, by using $M_{\rm P} = 1.22\times 10^{19}$ GeV, the
numerical values for the string scale are $m_s = 0.55$ GeV, 1820
TeV, and 385 PeV for $N$ = 2, 3, and 4, respectively.

\begin{figure}[b]
\begin{center}
\includegraphics[width=0.42\textwidth,angle=0]{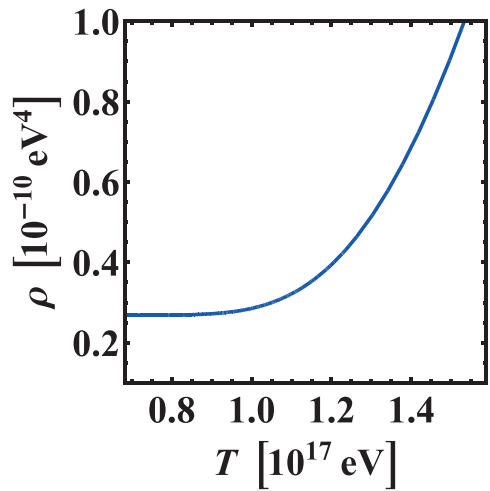}
\end{center}
\caption{Energy density ($\rho$) as a function of the temperature
(T), for $N=4$, $m_s=385$ PeV and small coupling $g_0\ll 1$.
\label{enef}}
\end{figure}

\begin{figure}[t]
\begin{center}
\includegraphics[width=0.4\textwidth,angle=0]{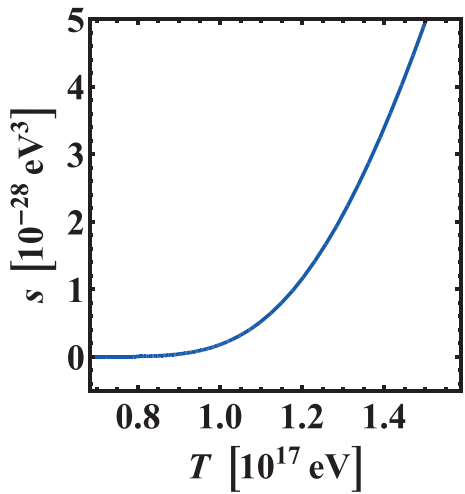}
\end{center}
\caption{Entropy density ($s$) as a function of the temperature (T),
for $N=4$, $m_s=385$ PeV and small coupling $g_0\ll 1$.
\label{enenf}}
\end{figure}

Rewriting the constraint (\ref{planck}) we have
\be g_o^2  = 2^7 (m_sR)^6\LF\frac{m_s}{ M_{\rm P}}\RF^2<
\LF\frac{m_s}{ M_{\rm P}}\RF^2\,. \ee
Thus we find that for $N=3,4$ the value of the string coupling can
at most be $10^{-15}$ and $10^{-12}$ respectively. For larger values
of $N$, the values of the string coupling can be larger. For
example, $N=9$ and $m_s=10^{15}$ GeV, allows us to have $g_o\sim
10^{-4}$. A coupling of order $g_o\sim 10^{-2}$ can be achieved  for
$N>11$, for instance with $N=12$, and $m_s=10^{17}$ GeV.   One can
see the typical dependence with temperature in Figure \ref{enef},
where we plot the energy density.  In Figure \ref{enenf}, one can
see the entropy density versus temperature.

Let us now look at the opposite case when $MR\gg 1$. In this case
\be \La=-\la (N-1)(2N-1)!!(2\pi R)^{6}\LF{M\over
2\sqrt{\pi}}\RF^{10N}\,. \ee
As compared to the $MR\ll1$ case, it is important to note that in
this case, $g_o$ and $R$ appears in different combinations in the
expression of $\La$ and $M_p$. This enables us to be able to choose
the value of $m_s$ freely. As before we can eliminate the
compactification radius in favor or $M_p$:
\be {\La\over M_p^4}=Q_{>}\LF{m_s\over M_p}\RF^{2}g_o^{2N}\with
Q_>\equiv {(N-1)(2N-1)!!\over 2^8N(\ln
(2N-1))^{5N}(2\pi)^{5N-6}}\LF{2(N-1)\over (2N-1)^2}\RF^{N-1}\,. \ee
Let us choose $m_s\sim$ 10 TeV. Ignoring order one factors, we find
the following estimates: $g_o \sim 10^{-22}, 10^{-14}$, $10^{-9}$,
and $10^{-2}$ for $p$ = 2, 3, 4, and 9 respectively. Again, as we
increase the values of $N$, the string coupling required for
producing the hierarchy becomes larger.

However, we should check that these values are consistent with
(\ref{planck}). It is easy to see that we must now satisfy \be g_o^2
> \LF\frac{m_s}{ M_{\rm P}}\RF^2\sim 10^{-30} \ee Thus we again find
that while the $N=2$ case is excluded, $N=3$ and higher are
consistent.

\section{Discussion and Conclusion}
\label{sec:conclusion}

We have analyzed the main thermodynamic properties of p-adic string models that describe the tachyon phenomenology of string theory. We have reproduced qualitatively many results of string theories at finite temperature already discussed in the literature; such as thermal duality at leading order for $p=3$, or the temperature dependence of radiative corrections.

On the other hand, p-adic models constitute a motivated example of
field theories characterized by non-local propagators. The finite
temperature computations we have developed constitute a basic
approach that may be generalized to other tachyonic field models
from string theory reduction, non-commutative field theories, or
quantum algebras. We do not mean that all the technical details and
specific results will apply, but rather the algorithms for the
perturbative treatment we developed should be. In any case, a
particular property of the p-adic case is the absent of poles in the
propagator. Non-local field theories with real states at finite
temperature are expected to have a contribution from the free theory
similar to standard thermodynamics.

One can contemplate non-local generalizations of the Standard Model.
From the field theory perspective, we have provided a unique
prescription of adding counter-terms loop-by-loop which ensures that
the theory remains ghost free and hence consistent. The
counter-terms are of fundamental importance since they determine the
thermodynamic properties of the model at low temperatures. In the
p-adic case, it is able to change the sign of the vacuum energy,
giving a positive cosmological constant that is suppressed by the
ratio of the string to the Planck scale. In fact, we are even able
to reproduce the observed value of the cosmological constant for a
range of relevant parameters. However, we need a small string
coupling to achieve this.

Our study unfolded as follows. First, we studied the finite temperature contribution of the free theory and determined that its vanishing contribution is consistent with the physical degrees of freedom of the theory. Specifically, the p-adic model gives a zero contribution due to the non existence of poles in the p-adic propagator.

The interacting theory is not trivial, and the first non-vanishing contribution to the partition function comes from the diagram with one vertex. For example, for $p=3$, this contribution is given by a 2-loop diagram that satisfies the well-known thermal duality of string theory, namely, a symmetry between the behaviors at low and high temperatures. However, we found that this result does not generalize to arbitrary $p$ and breaks down at higher order in the coupling, i.e. high temperatures. The symmetry is also broken if one takes into account the self-energy counter-term.

We then computed some of the typical higher order finite temperature loop corrections, such as ring, necklace and sunset diagrams. It is interesting to note that the non-local theories have a better UV behavior as compared to the local theories, and for the $p$-adic case, none of these individual diagrams are divergent. Also, we found that ring diagrams provide a series with good convergence properties, in contrast with the infrared divergences that are usually found in the local counter-parts.  On the other hand, the infinite series arising from the necklace diagrams diverges at a finite temperature. This behavior may be related to the fact that the $p$-adic potential is not bounded from below.

In general, the picture that has emerged from our study is that one
can trust perturbation theory as long as $T<m_s/g_o^2$. For small
string coupling this temperature can be quite high, and therefore we
can probe the Hagedorn transition.  However, once we reach $T\sim
m_s/g_o^2$, one can no longer trust perturbation theory, all the
higher order terms become important, and the partition function
seems to diverge.  In addition, at this temperature, contributions
from non-perturbative states start to be important
\cite{Biswas:2010yx}.  What happens at even higher temperatures is
an open question.

\section*
{\appendix{Appendix: Dimensional reduction of the vacuum
energy}}

In this appendix we perform the usual dimensional reduction
procedure, valid as long as $MR\ll 1$, to obtain the effective four
dimensional vacuum energy. We will see that this is identical to
what we derived in section \ref{sec:cc}. Let us start with the
higher dimensional $p$-adic action (\ref{action}):
$$
S =
\frac{\ms^D}{\bar{g}_p^2} \int d^D x  \left[ -\frac{1}{2} \varphi\,
{\rm e}^{-{\Box/ M^2}} \varphi+\frac{1}{2N} \varphi^{2N} \right]\,,
$$
where to maintain clarity we have introduced $\bar{g}_p$ to denote the $D$ dimensional coupling constant. When $MR\ll1$, one can perform the usual dimensional reduction where one only keeps the zero modes. This leads rather straight-forwardly to the effective four dimensional action
\be
S =
\frac{\ms^D(2\pi R)^{d}}{g_p^2} \int d^4 x  \left[ -\frac{1}{2}
\varphi\, {\rm e}^{-{\Box/ M^2}} \varphi+\frac{1}{2N} \varphi^{2N}
\right]\,, \ee where now the $\Box$ operator just represents the
four dimensional D'Alembertian. This is exactly of the same form as
the 4 dimensional $p$-adic action with the following identification
\be g_p^2={\bar{g}_p^2\over (2\pi R M_s)^{d}}\,. \ee
One can now simply read off the 2-loop four dimensional cosmological
constant from the result (\ref{pvacuum}) derived in section
\ref{sec:cc}:
%
%
%
\begin{eqnarray} \epsilon_{\rm vac} &=& -(N-1)(2N-1)!! \lambda \left(
\frac{M^2}{4\pi} \right)^{2N}\nonumber\\
&=&{(N-1)(2N-1)!!\over 2N} m_s^{-4(N-1)}\LT{\bar{g}_p^2\over (2\pi R
M_s)^{d}}\RT^{2(N-1)} \left( \frac{M^2}{4\pi} \right)^{2N}\,.
\end{eqnarray}
This is precisely the same result. Let us try to estimate this in
the general case of $d$ extra dimensions. For this purpose we note
that the dimensional reduction of the gravitational action leads to
a relation between the effective four-dimensional Planck's constant,
the string scale and the internal radius, of the form
\be M_p^2\sim{(2\pi R m_s)^d m_s^2\over g_o^2}\,. \ee
Rewriting the relation we find
\be {g_o^2\over (2\pi R m_s)^d}\sim \LF{m_s\over M_P}\RF^2\,. \ee
Now perimetrically, the vacuum energy is given by
\be \epsilon_{\rm vac}\sim m_s^4\LT{\bar{g}_o^2\over (2\pi R
M_s)^{d}}\RT^{2(N-1)}\sim m_s^4\LT{m_s\over M_p}\RT^{4(N-1)}\, , \ee
We can draw a very important conclusion from the above result,
namely that the hierarchial suppression of the vacuum energy in
$p$-adic models is independent of the number of extra dimensions, at
least in the limit of small radius, $MR\ll 1$.

This approach also allows us to estimate the effect of higher loops
on the value of the cosmological constant. As we argued in section
\ref{sec:higher-loops}, at low temperatures, higher loop corrections
are suppressed as $g_o^{l-1}$, where $g_o$ represents the effective
four-dimensional open string coupling. Thus, as long as we have
$g_o\ll 1$, we can trust our 2-loop result for the cosmological
constant. For the higher dimensional theory, this means \be
{\bar{g}_p^2\over (2\pi R M_s)^{d}}\sim { m_s^2\over M_P^2}\ll 1\,.
\ee To be specific, for the ten dimensional supergravity we have \be
{\bar{g}_p^2\over (2\pi R M_s)^{d}}\sim { 2m_s^2\over\pi^6 M_P^2}\ll
1\,. \ee Since $m_s<M_p$, this condition can easily be met.

\section*{Acknowledgements}

We thank Neil Barnaby, Gianluca Calcagni, Debashish Ghoshal and Bala
Sathiapalan  for useful and interesting conversations on the
subject. This work was supported by the U.S. DOE Grant Nos.
DE-FG02-87ER40328 and DOE/DE-FG02-94ER40823, the FPA 2008-00592
(DGICYT, Spain), the CAM/UCM 910309, and MICINN Consolider-Ingenio
MULTIDARK CSD2009-00064.


\begin{thebibliography}{99}

\bibitem{sft1}
 E.~Witten,
  ``Noncommutative Geometry and String Field Theory,''
  Nucl.\ Phys.\  B {\bf 268}, 253 (1986).

\bibitem{sft2}

  V.~A.~Kostelecky and S.~Samuel,
  ``The Static Tachyon Potential in the Open Bosonic String Theory,''
  Phys.\ Lett.\  B {\bf 207}, 169 (1988).

 V.~A.~Kostelecky and S.~Samuel,
  ``On a Nonperturbative Vacuum for the Open Bosonic String,''
  Nucl.\ Phys.\  B {\bf 336}, 263 (1990).

  I.~Y.~Aref'eva, A.~S.~Koshelev, D.~M.~Belov and P.~B.~Medvedev,
  ``Tachyon Condensation in Cubic Superstring Field Theory,''
  Nucl.\ Phys.\  B {\bf 638}, 3 (2002)
  [arXiv:hep-th/0011117].

  I.~Y.~Aref'eva, L.~V.~Joukovskaya and A.~S.~Koshelev,
  ``Time Evolution in Superstring Field Theory on Non-BPS Brane. I: Rolling
  Tachyon and Energy-Momentum Conservation,''
  JHEP {\bf 0309}, 012 (2003)
  [arXiv:hep-th/0301137].

  M.~Fujita and H.~Hata,
  ``Rolling Tachyon Solution in Vacuum String Field Theory,''
  Phys.\ Rev.\  D {\bf 70}, 086010 (2004)
  [arXiv:hep-th/0403031].

  T.~G.~Erler,
  ``Level Truncation and Rolling the Tachyon in the Lightcone Basis for Open
  String Field Theory,''
  arXiv:hep-th/0409179.

\bibitem{singular}


  G.~Calcagni and G.~Nardelli,
  ``Tachyon Solutions in Boundary and Cubic String Field Theory,''
  Phys.\ Rev.\  D {\bf 78}, 126010 (2008)
  [[arXiv:0708.0366] [hep-th]].

\bibitem{schnabl}

  M.~Schnabl,
  ``Analytic Solution for Tachyon Condensation in Open String Field Theory,''
  Adv.\ Theor.\ Math.\ Phys.\ {\bf 10}, 433 (2006)
  [arXiv:hep-th/0511286].

\bibitem{okawa}


  Y.~Okawa,
  ``Comments on Schnabl's Analytic Solution for Tachyon Condensation in
  Witten's Open String Field Theory,''
  JHEP {\bf 0604}, 055 (2006)
  [arXiv:hep-th/0603159].

\bibitem{analytic}

  M.~Kiermaier, Y.~Okawa, L.~Rastelli and B.~Zwiebach,
  ``Analytic Solutions for Marginal Deformations in Open String Field Theory,''
 JHEP {\bf 0801}, 028 (2008)
[arXiv:hep-th/0701249].

\bibitem{comments}

 M.~Schnabl,
  ``Comments on Marginal Deformations in Open String Field Theory,''
Phys.\ Lett.\  B {\bf 654}, 194 (2007).
[arXiv:hep-th/0701248].

\bibitem{taming}

  E.~Coletti, I.~Sigalov and W.~Taylor,
  ``Taming the Tachyon in Cubic String Field Theory,''
  JHEP {\bf 0508}, 104 (2005)
  [arXiv:hep-th/0505031].

\bibitem{ellwood}
 I.~Ellwood,
  ``Rolling to the Tachyon Vacuum in String Field Theory,''
JHEP {\bf 0712}, 028 (2007)
 [arXiv:0705.0013 [hep-th]].

\bibitem{pressure}

 N.~Jokela, M.~Jarvinen, E.~Keski-Vakkuri and J.~Majumder,
  ``Disk Partition Function and Oscillatory Rolling Tachyons,''
 J. Phys. A {\bf 41}, 015402 (2008)
[arXiv:0705.1916 [hep-th]].

\bibitem{sft_review}

 W.~Taylor and B.~Zwiebach,
  ``D-branes, Tachyons, and String Field Theory,''
  arXiv:hep-th/0311017.
\bibitem{claudia}

 C.~de Rham,
  ``The Effective Field Theory of Codimension-two Branes,''
 JHEP {\bf 0801}, 060 (2008)
[arXiv:0707.0884 [hep-th]].

\bibitem{padic_st}

  P.~G.~O.~Freund and M.~Olson,
  ``Nonarchimedean Strings,''
  Phys.\ Lett.\  B {\bf 199}, 186 (1987).

  P.~G.~O.~Freund and E.~Witten,
  ``Adelic String Amplitudes,''
  Phys.\ Lett.\  B {\bf 199}, 191 (1987).

 L.~Brekke, P.~G.~O.~Freund, M.~Olson and E.~Witten,
  ``Nonarchimedean String Dynamics,''
  Nucl.\ Phys.\  B {\bf 302}, 365 (1988).


\bibitem{zwiebach}
N.~Moeller and B.~Zwiebach,
  ``Dynamics with Infinitely Many Time Derivatives and Rolling Tachyons,''
  JHEP {\bf 0210}, 034 (2002)
  [arXiv:hep-th/0207107].

\bibitem{zeta}
  B.~Dragovich,
  ``Zeta Strings,''
  arXiv:hep-th/0703008.

B.~Dragovich,
  ``Zeta Nonlocal Scalar Fields,''
  Theor.\ Math.\ Phys.\  {\bf 157}, 1671 (2008)
  [arXiv:0804.4114 [hep-th]].

\bibitem{random}
M.~R.~Douglas and S.~H.~Shenker,
  ``Strings in Less than One-Dimension,''
  Nucl.\ Phys.\  B {\bf 335}, 635 (1990).

 D.~J.~Gross and A.~A.~Migdal,
  ``Nonperturbative Solution of the Ising Model on a Random Surface,''
  Phys.\ Rev.\ Lett.\  {\bf 64}, 717 (1990).

 E.~Brezin and V.~A.~Kazakov,
  ``Exactly Solvable Field Theories of Cosed Strings,''
  Phys.\ Lett.\  B {\bf 236}, 144 (1990).

\bibitem{marc}
 T.~Biswas, M.~Grisaru and W.~Siegel,
  ``Linear Regge Trajectories from Worldsheet Lattice Parton Field Theory,''
  Nucl.\ Phys.\  B {\bf 708}, 317 (2005)
  [arXiv:hep-th/0409089].
\bibitem{ghoshal}

  D.~Ghoshal,
  ``p-adic String Theories Provide Lattice Discretization to the Ordinary
  String Worldsheet,''
  Phys.\ Rev.\ Lett.\  {\bf 97}, 151601 (2006).
\bibitem{ncft}

  M.~R.~Douglas and N.~A.~Nekrasov,
  ``Noncommutative Field Theory,''
  Rev.\ Mod.\ Phys.\  {\bf 73}, 977 (2001)
  [arXiv:hep-th/0106048].

  R.~J.~Szabo,
  ``Quantum Field Theory on Noncommutative Spaces,''
  Phys.\ Rept.\  {\bf 378}, 207 (2003)
  [arXiv:hep-th/0109162].
\bibitem{minimal}

  S.~Hossenfelder,
  ``Self-Consistency in Theories with a Minimal Length,''
  Class.\ Quant.\ Grav.\  {\bf 23}, 1815 (2006)
  [arXiv:hep-th/0510245].

  S.~Hossenfelder,
  ``Interpretation of Quantum Field Theories with a Minimal Length Scale,''
  Phys.\ Rev.\  D {\bf 73}, 105013 (2006)
  [arXiv:hep-th/0603032].

  S.~Hossenfelder,
  ``A Note on Quantum Field Theories with a Minimal Length Scale,''
  Class. Quant. Grav. {\bf 25}, 038003 (2008)
[arXiv:0712.2811 [hep-th]].


\bibitem{kdv}

  A.~Ludu, R.~A.~Ionescu and W.~Greiner, ``Generalized KdV Equation for Fluid Dynamics and Quantum Algebras,'' Found. Phys. {\bf 26}, 665 (1996)
  [arXiv:q-alg/9612006].


\bibitem{fluid}


A.~Ludu and J.~P.~Draayer, ``Patterns on Liquid Surfaces: Cnoidal Waves, Compactons and Scaling,''
Physica D {\bf 123}, 82 (1998) [arXiv:physics/0003077].

\bibitem{futuret}
T. Biswas, in preparation.

\bibitem{math}
Y.~Volovich,
  ``Numerical Study of Nonlinear Equations with Infinite Number of
  Derivatives,''
  J.\ Phys.\ A {\bf 36}, 8685 (2003)
  [arXiv:math-ph/0301028].

V.~S.~Vladimirov and Y.~I.~Volovich,
  ``On the Nonlinear Dynamical Equation in the p-adic String Theory,''
  Theor.\ Math.\ Phys.\  {\bf 138}, 297 (2004)
  [Teor.\ Mat.\ Fiz.\  {\bf 138}, 355 (2004)] [arXiv:math-ph/0306018].

 V.~S.~Vladimirov,
  ``On the Equation of the p-adic Open String for the Scalar Tachyon Field,''
  arXiv:math-ph/0507018.

D.~V.~Prokhorenko,
  ``On Some Bonlinear Integral Equation in the (Super)String Theory,''
  arXiv:math-ph/0611068.


\bibitem{neil}
 N.~Barnaby and N.~Kamran,
  ``Dynamics with Infinitely Many Derivatives: Variable Coefficient
  JHEP {\bf 0812}, 022 (2008)
  [arXiv:0809.4513 [hep-th]].

\bibitem{gianluca}

  G.~Calcagni, M.~Montobbio and G.~Nardelli,
  ``Localization of Nonlocal Theories,''
  Phys.\ Lett.\  B {\bf 662}, 285 (2008)
  [arXiv:0712.2237 [hep-th]].


\bibitem{bbc}

N.~Barnaby, T.~Biswas and J.~M.~Cline,
  ``p-adic Inflation,''
  JHEP {\bf 0704}, 056 (2007)
  [arXiv:hep-th/0612230].

J.~E.~Lidsey,
  ``Stretching the Inflaton Potential with Kinetic Energy,''
  Phys.\ Rev.\  D {\bf 76}, 043511 (2007)
  [arXiv:hep-th/0703007].

 N.~J.~Nunes and D.~J.~Mulryne,
  ``Non-linear Non-local Cosmology,''
  AIP Conf.\ Proc.\  {\bf 1115}, 329 (2009)
  [arXiv:0810.5471 [astro-ph]].

N.~Barnaby and J.~M.~Cline,
  ``Large Nongaussianity from Nonlocal Inflation,''
  JCAP {\bf 0707}, 017 (2007)
  [arXiv:0704.3426 [hep-th]].

N.~Barnaby and J.~M.~Cline,
  ``Predictions for Nongaussianity from Nonlocal Inflation,''
  JCAP {\bf 0806}, 030 (2008)
  [arXiv:0802.3218 [hep-th]].

\bibitem{cosmo}
I.~Y.~Aref'eva and L.~V.~Joukovskaya,
  ``Time Lumps in Nonlocal Stringy Models and Cosmological Applications,''
  JHEP {\bf 0510}, 087 (2005)
  [arXiv:hep-th/0504200].

I.~Y.~Aref'eva, A.~S.~Koshelev and S.~Y.~Vernov,
  ``Crossing of the w=-1 Barrier by D3-brane Dark Energy Model,''
 Phys.\ Rev.\  D {\bf 72}, 064017 (2005)
  [arXiv:astro-ph/0507067].


\bibitem{bouncing}
 I.~Y.~Aref'eva, L.~V.~Joukovskaya and S.~Y.~Vernov,
  ``Bouncing and Accelerating Solutions in Nonlocal Stringy Models,''
  JHEP {\bf 0707}, 087 (2007)
  [arXiv:hep-th/0701184].


\bibitem{warren}
T.~Biswas, A.~Mazumdar and W.~Siegel,
  ``Bouncing Universes in String-Inspired Gravity,''
  JCAP {\bf 0603}, 009 (2006)
  [arXiv:hep-th/0508194].

T.~Biswas, T.~Koivisto and A.~Mazumdar,
``Resolution of the Big Crunch/Bang Singularity in Non-local Higher Derivative
theories of Gravity,'' in preparation.

\bibitem{cosmology}
G.~Calcagni and G.~Nardelli,
  ``Cosmological Rolling Solutions of Nonlocal Theories,''
Int. J. Mod. Phys. D {\bf 19}, 329 (2010)
[arXiv:0904.4245 [hep-th]].

G.~Calcagni and G.~Nardelli,
  ``Nonlocal Instantons and Solitons in String Models,''
  Phys.\ Lett.\  B {\bf 669}, 102 (2008)
  [arXiv:0802.4395 [hep-th]].

G.~Calcagni, M.~Montobbio and G.~Nardelli,
  ``Route to Nonlocal Cosmology,''
  Phys.\ Rev.\  D {\bf 76}, 126001 (2007)
  [arXiv:0705.3043 [hep-th]].

\bibitem{minahan}
  J.~A.~Minahan,
  ``Quantum Corrections in p-adic String Theory,''
  arXiv:hep-th/0105312.

\bibitem{prl-old}
  T.~Biswas, J.~A.~R.~Cembranos and J.~I.~Kapusta,
 ``Thermal Duality and Hagedorn Transition from p-adic Strings,''
  Phys.\ Rev.\ Lett.\  {\bf 104}, 021601 (2010)
  [arXiv:0910.2274 [hep-th]].

\bibitem{seiberg}
  J.~L.~Davis, F.~Larsen and N.~Seiberg,
  ``Heterotic Strings in Two Dimensions and New Stringy Phase Transitions,''
  JHEP {\bf 0508}, 035 (2005)
  [arXiv:hep-th/0505081].

  N.~Seiberg,
  ``Long Strings, Anomaly Cancellation, Phase Transitions, T-duality and
  Locality in the 2d Heterotic String,''
  JHEP {\bf 0601}, 057 (2006)
  [arXiv:hep-th/0511220].

 J.~L.~Davis,
  ``The Moduli Space and Phase Structure of Heterotic Strings in Two
  Dimensions,''
  Phys.\ Rev.\  D {\bf 74}, 026004 (2006)
  [arXiv:hep-th/0511298].

\bibitem{vafa}
 A.~A.~Tseytlin and C.~Vafa,
  ``Elements Of String Cosmology,''
  Nucl.\ Phys.\  B {\bf 372}, 443 (1992)
  [arXiv:hep-th/9109048].

\bibitem{robert}
R.~H.~Brandenberger and C.~Vafa,
  ``Superstrings in the Early Universe,''
  Nucl.\ Phys.\  B {\bf 316}, 391 (1989).

\bibitem{hagedorn-cosmology}
S.~Alexander and T.~Biswas,
  ``The Cosmological BCS mechanism and the Big Bang Singularity,''
  Phys.\ Rev.\  D {\bf 80}, 023501 (2009)
  [arXiv:0807.4468 [hep-th]].

T.~Biswas,
  ``The Hagedorn Soup and an Emergent Cyclic Universe,''
  arXiv:0801.1315 [hep-th].



B.~Greene, D.~Kabat and S.~Marnerides,
  ``Bouncing and Cyclic String Gas Cosmologies,''
  Phys.\ Rev.\  D {\bf 80}, 063526 (2009)
  [arXiv:0809.1704 [hep-th]].

 R.~Danos, A.~R.~Frey and A.~Mazumdar,
  ``Interaction Rates in String Gas Cosmology,''
  Phys.\ Rev.\  D {\bf 70}, 106010 (2004)
  [arXiv:hep-th/0409162].

 D.~Skliros and M.~Hindmarsh,
  ``Large Radius Hagedorn Regime in String Gas Cosmology,''
Phys. Rev. D {\bf 78}, 063539 (2008)
[arXiv:0712.1254 [hep-th]].

\bibitem{thermal}
 A.~Nayeri, R.~H.~Brandenberger and C.~Vafa,
  ``Producing a Scale-Invariant Spectrum of Perturbations in a Hagedorn Phase
  of String Cosmology,''
  Phys.\ Rev.\ Lett.\  {\bf 97}, 021302 (2006)
  [arXiv:hep-th/0511140].

  R.~H.~Brandenberger, A.~Nayeri, S.~P.~Patil and C.~Vafa,
  ``String Gas Cosmology and Structure Formation,''
  Int.\ J.\ Mod.\ Phys.\  A {\bf 22}, 3621 (2007)
  [arXiv:hep-th/0608121].

  R.~H.~Brandenberger,
  ``String Gas Cosmology and Structure Formation: A Brief Review,''
  Mod.\ Phys.\ Lett.\  A {\bf 22}, 1875 (2007)
  [arXiv:hep-th/0702001].

\bibitem{hag-bounce}
 T.~Biswas, R.~Brandenberger, A.~Mazumdar and W.~Siegel,
  JCAP {\bf 0712}, 011 (2007)
  [arXiv:hep-th/0610274].

\bibitem{lee-wick}
 B.~Grinstein, D.~O'Connell and M.~B.~Wise,
  ``The Lee-Wick Standard Model,''
  Phys.\ Rev.\  D {\bf 77}, 025012 (2008)
  [arXiv:0704.1845 [hep-ph]].

  T.~D.~Lee and G.~C.~Wick,
  ``Finite Theory of Quantum Electrodynamics,''
  Phys.\ Rev.\  D {\bf 2}, 1033 (1970).

\bibitem{moffat}
  J.~W.~Moffat and V.~T.~Toth,
  ``Redesigning Electroweak Theory: Does the Higgs Particle Exist?,''
  arXiv:0908.0780 [hep-ph].

 J.~W.~Moffat and V.~T.~Toth,
  ``A Finite Electroweak Model without a Higgs Particle,''
  arXiv:0812.1991 [hep-ph].

J.~W.~Moffat,
  ``Electroweak Model Without A Higgs Particle,''
  arXiv:0709.4269 [hep-ph].

\bibitem{witten}
P.~G.~O.~Freund and M.~Olson,
  ``Nonarchimedean Strings,''
  Phys.\ Lett.\  B {\bf 199}, 186 (1987).

P.~G.~O.~Freund and E.~Witten,
  ``Adelic String Amplitudes,''
  Phys.\ Lett.\  B {\bf 199}, 191 (1987).

 L.~Brekke, P.~G.~O.~Freund, M.~Olson and E.~Witten,
  ``Nonarchimedean String Dynamics,''
  Nucl.\ Phys.\  B {\bf 302}, 365 (1988).

\bibitem{Frampton}
 P.~H.~Frampton and Y.~Okada,
   ``The p-adic String N Point Function,''
  Phys.\ Rev.\ Lett.\  {\bf 60}, 484 (1988).

P.~H.~Frampton and Y.~Okada,
``Effective Scalar Field Theory of p-adic String,''
  Phys.\ Rev.\  D {\bf 37}, 3077 (1988).

\bibitem{p=1}
A.~A.~Gerasimov and S.~L.~Shatashvili,
  ``On Exact Tachyon Potential in Open String Field Theory,''
  JHEP {\bf 0010}, 034 (2000)
  [arXiv:hep-th/0009103].

\bibitem{Sen:p-adic}
 D.~Ghoshal and A.~Sen,
  ``Tachyon Condensation and Brane Descent Relations in p-adic String
  Theory,''
  Nucl.\ Phys.\  B {\bf 584}, 300 (2000)
  [arXiv:hep-th/0003278].

\bibitem{Minahan:ModeInteractions}
 J.~A.~Minahan,
  ``Mode Interactions of the Tachyon Condensate in p-adic String Theory,''
  JHEP {\bf 0103}, 028 (2001)
  [arXiv:hep-th/0102071].

\cite{Biswas:2010yx}
\bibitem{Biswas:2010yx}
  T.~Biswas, J.~A.~R.~Cembranos and J.~I.~Kapusta,
  arXiv:1006.4098 [hep-th].


 \bibitem{KapGale}
J. I. Kapusta and C. Gale, {\it Finite Temperature Field Theory}, Cambridge University Press, Cambridge, 2nd edition, 2006.


\bibitem{bala-old}
B.~Sathiapalan,
  ``Vortices on the String World Sheet and Constraints on Toral
  Compactification,''
  Phys.\ Rev.\  D {\bf 35}, 3277 (1987).

 Y.~I.~Kogan,
  ``Vortices On The World Sheet And String's Critical Dynamics,''
  JETP Lett.\  {\bf 45}, 709 (1987)
  [Pisma Zh.\ Eksp.\ Teor.\ Fiz.\  {\bf 45}, 556 (1987)].

\bibitem{atick}
J.~J.~Atick and E.~Witten,
  ``The Hagedorn Transition and the Number of Degrees of Freedom of String
  Theory,''
  Nucl.\ Phys.\  B {\bf 310}, 291 (1988).

\bibitem{Polchinski:1985zf}




 J.~Polchinski,
  ``Evaluation Of The One Loop String Path Integral,''
  Commun.\ Math.\ Phys.\  {\bf 104}, 37 (1986).

\bibitem{bala}
 B.~Sathiapalan and N.~Sircar,
  ``Can the Hagedorn Phase Transition be Explained from Matrix Model for
  Strings?,''
  JHEP {\bf 0808}, 019 (2008)
  [arXiv:0805.0076 [hep-th]].


\bibitem{dienes}
  K.~R.~Dienes and M.~Lennek,
  ``Re-identifying the Hagedorn Transition,''
  arXiv:hep-th/0505233.

\bibitem{jain}
  N.~Deo, S.~Jain, O.~Narayan and C.~I.~Tan,
  ``The Effect of Topology on the Thermodynamic Limit for a String Gas,''
  Phys.\ Rev.\  D {\bf 45}, 3641 (1992).

  N.~Deo, S.~Jain and C.~I.~Tan,
  ``String Statistical Mechanics above Hagedorn Energy Density,''
  Phys.\ Rev.\  D {\bf 40}, 2626 (1989).

  N.~Deo, S.~Jain and C.~I.~Tan,
  ``Strings at High-Energy Densities and Complex Temperature,''
  Phys.\ Lett.\  B {\bf 220}, 125 (1989).

\bibitem{chaudhuri}
S.~Chaudhuri, ``Dispelling the Hagedorn Myth: Canonical and Microcanonical Strings,'' arXiv:hep-th/0506143.

\bibitem{smilga}
  A.~V.~Smilga,
  ``Ghost-Free Higher-Derivative Theory,''
  Phys.\ Lett.\  B {\bf 632}, 433 (2006)
  [arXiv:hep-th/0503213].


\end{thebibliography}
\end{document}